\documentclass[12pt]{article}
\usepackage{amsmath,verbatim,amssymb,epsfig,rotating,longtable,xcolor}
\usepackage{setspace}
\usepackage{color, colortbl}
\usepackage{multirow}
\usepackage{enumitem}
\newcommand{\sspace}{\fontsize{12}{8}\selectfont} 
\usepackage{lipsum}
\setlist[description]{leftmargin=\parindent,labelindent=\parindent}
\definecolor{Gray}{gray}{0.9}
\begin{document}
	\def\eqx"#1"{{\label{#1}}}
	\def\eqn"#1"{{\ref{#1}}}
	
	\makeatletter 
	\@addtoreset{equation}{section}
	\makeatother  
	
	\def\yincomment#1{\vskip 2mm\boxit{\vskip 2mm{\color{red}\bf#1} {\color{blue}\bf --Yin\vskip 2mm}}\vskip 2mm}
	\def\lincomment#1{\vskip 2mm\boxit{\vskip 2mm{\color{blue}\bf#1} {\color{black}\bf --Lin\vskip 2mm}}\vskip 2mm}
	\def\squarebox#1{\hbox to #1{\hfill\vbox to #1{\vfill}}}
	\def\boxit#1{\vbox{\hrule\hbox{\vrule\kern6pt
				\vbox{\kern6pt#1\kern6pt}\kern6pt\vrule}\hrule}}

	\def\theequation{\thesection.\arabic{equation}}
	\newcommand{\ds}{\displaystyle}
	
	\newcommand{\bJ}{\mbox{\bf J}}
	\newcommand{\bF}{\mbox{\bf F}}
	\newcommand{\bM}{\mbox{\bf M}}
	\newcommand{\bR}{\mbox{\bf R}}
	\newcommand{\bZ}{\mbox{\bf Z}}
	\newcommand{\bX}{\mbox{\bf X}}
	\newcommand{\bx}{\mbox{\bf x}}
	\newcommand{\bww}{\mbox{\bf w}}
	\newcommand{\bQ}{\mbox{\bf Q}}
	\newcommand{\bH}{\mbox{\bf H}}
	\newcommand{\bh}{\mbox{\bf h}}
	\newcommand{\bz}{\mbox{\bf z}}
	\newcommand{\br}{\mbox{\bf r}}
	\newcommand{\ba}{\mbox{\bf a}}
	\newcommand{\be}{\mbox{\bf e}}
	\newcommand{\bG}{\mbox{\bf G}}
	\newcommand{\bB}{\mbox{\bf B}}
	\newcommand{\bb}{\mbox{\bf b}}
	\newcommand{\bA}{\mbox{\bf A}}
	\newcommand{\bC}{\mbox{\bf C}}
	\newcommand{\bI}{\mbox{\bf I}}
	\newcommand{\bD}{\mbox{\bf D}}
	\newcommand{\bU}{\mbox{\bf U}}
	\newcommand{\bc}{\mbox{\bf c}}
	\newcommand{\bd}{\mbox{\bf d}}
	\newcommand{\bs}{\mbox{\bf s}}
	\newcommand{\bS}{\mbox{\bf S}}
	\newcommand{\bV}{\mbox{\bf V}}
	\newcommand{\bv}{\mbox{\bf v}}
	\newcommand{\bW}{\mbox{\bf W}}
	\newcommand{\bY}{\mathbf{ Y}}
	\newcommand{\bw}{\mbox{\bf w}}
	\newcommand{\bg}{\mbox{\bf g}}
	\newcommand{\bu}{\mbox{\bf u}}
	\newcommand{\mI}{\mbox{I}}
	
	\newcommand{\bch}{\color{blue}\it}
	\newcommand{\ech}{\color{black}\rm}
	
	\def\bb{{\bf b}}
	
	\newcommand{\bcU}{\boldsymbol{\cal U}}
	\newcommand{\bbeta}{\boldsymbol{\beta}}
	\newcommand{\bdelta}{\boldsymbol{\delta}}
	\newcommand{\bDelta}{\boldsymbol{\Delta}}
	\newcommand{\boldeta}{\boldsymbol{\eta}}
	\newcommand{\bxi}{\boldsymbol{\xi}}
	\newcommand{\bGamma}{\boldsymbol{\Gamma}}
	\newcommand{\bSigma}{\boldsymbol{\Sigma}}
	\newcommand{\balpha}{\boldsymbol{\alpha}}
	\newcommand{\bOmega}{\boldsymbol{\Omega}}
	\newcommand{\btheta}{\boldsymbol{\theta}}
	\newcommand{\bepsilon}{\boldsymbol{\epsilon}}
	\newcommand{\bmu}{\boldsymbol{\mu}}
	\newcommand{\bnu}{\boldsymbol{\nu}}
	\newcommand{\bgamma}{\boldsymbol{\gamma}}
	\newcommand{\btau}{\boldsymbol{\tau}}
	\newcommand{\bTheta}{\boldsymbol{\Theta}}

	\newtheorem{thm}{Theorem}
	\newtheorem{lem}{Lemma}[section]
	\newtheorem{rem}{Remark}[section]
	\newtheorem{cor}{Corollary}[section]

	\newcommand{\tabincell}[2]{\begin{tabular}{@{}#1@{}}#2\end{tabular}}

\def\eqx"#1"{{\label{#1}}}
\def\eqn"#1"{{\ref{#1}}}

\makeatletter 
\@addtoreset{equation}{section}
\makeatother  

\def\yincomment#1{\vskip 2mm\boxit{\vskip 2mm{\color{red}\bf#1} {\color{blue}\bf --Yin\vskip 2mm}}\vskip 2mm}
\def\lincomment#1{\vskip 2mm\boxit{\vskip 2mm{\color{blue}\bf#1} {\color{black}\bf --Lin\vskip 2mm}}\vskip 2mm}
\def\squarebox#1{\hbox to #1{\hfill\vbox to #1{\vfill}}}
\def\boxit#1{\vbox{\hrule\hbox{\vrule\kern6pt
			\vbox{\kern6pt#1\kern6pt}\kern6pt\vrule}\hrule}}

\baselineskip=24pt
\begin{center}
	{\Large \bf  Incorporating historical information and real-world evidence to improve phase I clinical trials}
\end{center}

\begin{center}
	{\bf Yanhong Zhou$^1$, J.Jack Lee$^1$,  Shunguang Wang$^2$, Stuart Bailey$^2$ and Ying Yuan$^{1}$}
\end{center}
\begin{center}

	$^1$Department of Biostatistics, The University of Texas MD Anderson Cancer Center\\
	Houston, TX\\
	$^2$ Novartis Institutes for BioMedical Research, Cambridge, MA \\

	\vspace{2mm}

\noindent{\textbf{Abstract}}\end{center}

Incorporating historical data or real-world evidence has a great potential to improve the efficiency of phase I clinical trials and to accelerate drug development. For model-based  designs, such as the continuous reassessment method (CRM), this can be conveniently carried out by specifying a ``skeleton," i.e., the prior estimate of dose limiting toxicity (DLT) probability at each dose. In contrast, little work has been done to incorporate historical data or real-world evidence into model-assisted designs, such as the Bayesian optimal interval (BOIN), keyboard, and modified toxicity probability interval (mTPI) designs. This has led to the misconception that model-assisted designs cannot incorporate prior information. In this paper, we propose a unified framework that allows for incorporating historical data or real-world evidence into model-assisted designs. The proposed approach uses the well-established ``skeleton" approach, combined with the concept of prior effective sample size, thus it is easy to understand and use. More importantly, our approach maintains the hallmark of model-assisted designs: simplicity---the dose escalation/de-escalation rule can be tabulated prior to the trial conduct. Extensive simulation studies show that the proposed method can effectively incorporate prior information to improve the operating characteristics of model-assisted designs, similarly to model-based designs.
\vspace{.25cm}


\section{Introduction}
Recently, there has been tremendous interest in the use of prior information, such as historical data or real-world evidence, as an effective approach to improve the efficiency of clinical trials. In May 2019, the Food and Drug Administration (FDA) released a draft of guidelines for submitting documents using real-world data or evidence to the FDA for drugs and biologics \cite{us2019submitting}. When designing phase I trials, prior information is often available from previous studies.  For example, the drug to be investigated has been studied previously in other indications, or similar drugs belonging to the same class have been studied in earlier phase I trials \cite{zohar2011approach}. Another example is the proposal of bridging phase I trials to extend a drug from one ethnic group (e.g., Caucasian) to another (e.g., Asian) \cite{liu2015bridging}, or from adult patients to pediatric patients \cite{petit2018unified}. In this case, dose-toxicity data from the original trial in one ethnic group or adult can be used to inform the design of the subsequent bridging trials, see for example Liu et al. \cite{liu2015bridging},  Morita \cite{morita2011application},  and Li and Yuan \cite{liyuan2019}.

Various phase I trial designs have been proposed to find the maximum tolerated dose (MTD). These designs can be classified into algorithm-based, model-based, and model-assisted designs, depending on their statistical foundations and implementation approaches. Algorithm-based designs, such as the 3+3 design, are ad-hoc and simple to implement, but also rigid, with poor accuracy to identify the MTD. It is difficult, if not impossible, to incorporate prior information into algorithm-based designs. Model-based designs assume a dose-toxicity model and determine the dose escalation/de-escalation by continuously updating the estimate of the model based on accrued data. Typical examples of model-based designs are the continual reassessment method (CRM \cite{o1990continual}) and its variations, such as the escalation with overdose control \cite{babb1998cancer}, the Bayesian logistic regression model \cite{neuenschwander2008critical},  and the Bayesian model averaging CRM \cite{yin2009bayesian}. Model-based designs yield better performance than the 3+3 design in identifying and allocating more patients to MTD. Another important strength of model-based designs is that they are straightforward to incorporate prior information. In particular, for CRM, prior information can be easily incorporated by specifying a ``skeleton" of the dose-toxicity model---the prior estimate of dose limiting toxicity (DLT) probability for each dose. More details are provided later. Along that line, Liu et al.  proposed to bridge CRM for phase I clinical trials in different ethnic populations based on Bayesian model averaging \cite{liu2015bridging};  Morita proposed to incorporate informative prior to CRM \cite{morita2011application} ; Li and Yuan proposed the continuous reassessment method for pediatric phase I oncology trials (PA-CRM) to leverage trial information from adult trials to pediatric trials  \cite{liyuan2019}.

Model-assisted designs were developed to combine the simplicity of algorithm-based designs with the superior performance of model-based designs. Similar to model-based designs, model-assisted designs use a statistical model (e.g., the binomial model) to derive decision rules for efficient decision making. And like algorithm-based designs, model-assisted designs can have their dose escalation and de-escalation rules determined before the onset of a trial, and thus can be implemented as simply as algorithm-based designs. Examples of model-assisted designs include the Bayesian optimal interval (BOIN) design \cite{liu2015bayesian}, the modified toxicity probability interval design (mTPI \cite{ji2010modified}), and the keyboard design \cite{yan2017keyboard} (or mTPI-2 \cite{guo2017bayesian}). Extensive numerical studies show that the model-assisted designs yield superior performance comparable to more complicated model-based designs, and they are increasingly used in practice \cite{yuan2019model}. 

Model-assisted designs were developed assuming a non-informative prior. Little research has been done on how to incorporate informative prior information into the derivation of these designs. This has led to the misconception that model-assisted designs cannot incorporate informative prior information, which is sometimes cited as their weakness compared to model-based designs.

In this paper, we propose a unified framework to incorporate informative prior information into model-assisted designs, including BOIN and keyboard (or equivalently, mTPI-2) designs. Our method uses the skeleton approach similar to that in CRM, combined with the concept of prior effective sample size (PESS) \cite{morita2008determining}. The method is intuitive and easy to understand; more importantly, it maintains the simplicity of the model-assisted designs in the sense that their dose escalation/de-escalation rule can still be determined and included in the protocol before the onset of a trial. Numerical studies show that incorporating appropriate informative prior information can improve the performance of the model-assisted designs, similarly to CRM.

The remainder of the paper is organized as follows. In Section \ref{sec:method}, we propose the methodology of incorporating informative prior information through skeleton and PESS for CRM, BOIN, and keyboard designs. In Section \ref{sec:app}, we provide the software to implement the proposed designs.  In Section \ref{sec:simulation}, we conduct extensive simulation to evaluate the operating characteristics of the proposed methodology.  We conclude the study with a brief summary in Section \ref{sec:discussion}.

\section{Method}\label{sec:method}
\subsection{Incorporate prior information in CRM}
We first describe how to incorporate prior information in the CRM. Let $j=1, \cdots, J$, denote the $J$ doses under investigation, and $p_j$ denote the true DLT probability of dose $j$. The objective of the trial is to find the MTD, whose DLT probability is equal or the closest to a prespecified target DLT probability $\phi$. 

To incorporate prior information on the dose-toxicity relationship, we elicit the prior estimate of $(p_1, \cdots, p_J)$, denoted as $(q_1, \cdots, q_J)$, known as ``skeleton." The skeleton can be estimated based on historical data or real-world evidence, e.g., by fitting a logistic model or nonparametric model \cite{liu2015bridging}, or specified by clinicians based on their clinical experience. We link $p_j$ with the skeleton through a parametric model
\begin{equation}\label{crmmodel}
p_j =q_j^{{\rm exp}(\alpha)}, \quad \mbox{for} \quad j=1, \cdots, J,
\end{equation}
where $\alpha$ is an unknown parameter that controls the discrepancy between the prior estimate $q_j$ and the true DLT probability $p_j$.  Under the Bayesian paradigm,  we assign $\alpha$  a normal prior $f(\alpha) = N(0, \sigma^2)$, where $\sigma^2$ is a prespecified hyperparameter. As a result, {\it a priori} the dose-toxicity curve $(p_1, \cdots, p_J)$ centers around the skeleton $(q_1, \cdots, q_J)$. The value of $\sigma^2$ controls the amount of information contained in the prior that can be borrowed from historical data (i.e., the skeleton). A smaller value leads to stronger borrowing. If $\sigma^2=0$, the prior completely dominates the observed data, and $p_j \equiv q_j$ regardless of the observed data.

Let $D=(D_1, \cdots, D_J)$ denote the observed data, where $D_j=(n_j, y_j)$ denotes the data observed at dose level $j$ with $n_j$ being the number of patients treated and $y_j$ the number of patients who experienced DLTs at dose $j$. To make the decision of dose escalation and de-escalation, CRM updates the posterior estimate of $p_j$ as follows:
\begin{align*}
\hat{p}_j=\int q_j ^{{\rm exp}(\alpha)}\frac{L( D\mid \alpha) f(\alpha)}{\int L(D\mid \alpha) f(\alpha)d\alpha}d\alpha,	
\end{align*}
where $
L(D\mid \alpha) =\prod_{j=1}^{J}\left\{q_j^{ \exp(\alpha)}\right\}^{y_j}\left\{1-q_j^{\exp (\alpha)}\right\}^{n_j-y_j}
$
is the likelihood function, and $f(\alpha)=N(0, \sigma^2)$ is the prior distribution of $\alpha$. Then, CRM assigns the next cohort of patients at the dose whose $\hat{p}_j$ is closet to $\phi$. In practice, we typically impose safety rules, such as starting at the lowest dose level and no dose skipping during dose escalation.

Given a specific skeleton and prior $f(\alpha) = N(0, \sigma^2)$, it is of great importance to quantify how much information is borrowed from historical data. This is, however, rarely discussed in the dose finding literature. In what follows, we propose a simple and intuitive approach to formally quantify the information borrowed through the skeleton using the concept of prior effective sample size (PESS), which represents the sample size that the prior information is equivalent to. Morita et al. \cite{morita2008determining} proposed a general methodology to determine PESS, but this approach requires complicated derivation and intensive simulation.

Our approach is simpler, more intuitive, and built upon the following observation: assuming $y_j$ follows a binomial distribution $Binom(n_j, p_j)$, if $p_j$ follows a beta prior distribution $Beta(a, b)$, then $a+b$ can be interpreted as the PESS. Our strategy is to approximate the prior distribution of $p_j$, induced by model (\ref{crmmodel}) and $f(\alpha)$, with a beta distribution by matching the first and second moments. Therefore, PESS can be easily determined.  Specifically, given skeleton $(q_1, \cdots, q_J)$ and prior $f(\alpha)$, let $\mu_j$ and $\tau_j^2$ denote the  prior mean and variance of $p_j$, respectively, with

$$
\mu_j = \int p_j f(p_j) {\rm d} p_j,  \qquad \tau_j^2 = \int p_j^2 f(p_j){\rm d} p_j
-u_j^2, $$
where $f(p_j)$ is the prior distribution of $p_j$ induced by the prior distribution of $f(\alpha) =  N(0, \sigma^2)$, given by
$$
f(p_j)=-\frac{1}{\sqrt{2\pi}\sigma}\exp\left\{-\frac{\left[\log\left(\frac{\log(p_j)}{\log(q_j)}\right)\right]^2}{2\sigma^2}\right\}\frac{1}{p_j \log(p_j)}.
$$
 Matching the first and second comments of $p_j$ by a beta distribution $Beta(a_j, b_j)$, we obtain the skeleton PESS as $a_j+b_j$, where
\begin{align}
a_j= \frac{\mu_j^2(1-\mu_j)}{{\tau}^2_j}-\mu_j,
\,\, \,\,b_j=\frac{a (1-\mu_j)}{\mu_j}.
\end{align}

This reveals a property of CRM that is barely discussed in the literature, though it is of great importance in practice. Because $p_j$ is a non-linear function of $\alpha$, once prior $f(\alpha)$ is specified, PESS for each dose is automatically determined.  For example, given skeleton $(q_1, \cdots, q_5) = (0.10, 0.19, 0.30, 0.42, 0.54)$ and prior  $f(\alpha)=N(0, 0.72)$, PESS is (3, 3, 3, 3.1, 3.4) for the five doses. As a result, CRM does not allow users to specify dose-specific prior information or PESS. However, in practice, we often have an unequal amount of prior information for different doses.  For example, we often have more data at the doses that are below and around the MTD from historical phase I trials. In this case, it is highly desirable to be able to specify a different PESS for each unique dose according to the historical data.

\subsection{Incorporate prior information in BOIN}
We now discuss how to use the skeleton, coupled with PESS, to incorporate prior information into model-assisted designs such as BOIN. To do so, we first briefly describe the genesis of the BOIN design, which lays the foundation for the proposed approach.
Consider a class of nonparametric designs ${\cal C}_{np}$ as follows.
\vspace{2mm}
\begin{itemize}
	\item[(a)] Patients in the first cohort are treated at the lowest or a prespecified starting dose level.
	
	\item[(b)] At the current dose level $j$, let ${\hat p_j} = {{{y_j}} \mathord{\left/ {\vphantom {{{m_j}} {{n_j}}}} \right. \kern-\nulldelimiterspace} {{n_j}}}$ denote the observed DLT probability, and ${\lambda _{e}}\left( {j, {n_j},\phi } \right)$ and ${\lambda _{d}}\left( {j, {n_j},\phi } \right)$ denote arbitrary functions of $j$, $n_j$ and $\phi$, serving as the dose escalation and de-escalation boundaries, respectively, with $0 \le {\lambda _{e}}\left( {j, {n_j},\phi } \right) < {\lambda _{d}}\left( {j, {n_j},\phi } \right) \le 1$. Use the following procedure to assign a dose to the next cohort of patients. 
	\begin{itemize}
		\item[$\bullet$] Escalate the dose level to $j$ + 1, if ${\hat p_j} < {\lambda _{e}}\left( {{j, n_j},\phi } \right)$;
		
		\item[$\bullet$] De-escalate the dose level to $j$ - 1, if ${\hat p_j} > {\lambda _{d}}\left( {{j, n_j},\phi } \right)$;
		
		\item[$\bullet$] Stay at the same dose level, $j$, if ${\lambda _{e}}\left( {{j, n_j},\phi } \right) \le {\hat p_j} \le {\lambda _{d}}\left( {{j, n_j},\phi } \right)$.
	\end{itemize}

	\item[(c)] This process is continued until the maximum sample size is reached.
	
\end{itemize}
Note that ${\lambda _{e}}\left( {j, {n_j},\phi } \right)$ and ${\lambda _{d}}\left( {j, {n_j},\phi } \right)$ can vary with dose level $j$, the number of patients treated $n_j$, and the target $\phi$. This class of nonparametric designs includes all possible designs that do not impose a parametric assumption on the dose-toxicity curve. For notational brevity, in what follows, we suppress arguments in ${\lambda _{e}}\left( {j, {n_j},\phi } \right)$ and ${\lambda _{d}}\left( {j, {n_j},\phi } \right)$ and denote them as $\lambda_e$ and $\lambda_d$.

The BOIN design is obtained by choosing the optimal dose escalation and de-escalation boundaries ${\lambda _{e}}$ and ${\lambda _{d}}$ to minimize the probability of making incorrect dose escalation and de-escalation decisions.  The optimization is carried out under three point hypotheses:
$$H_1: p_j = \phi; \quad H_2: p_j = \phi_1; \quad H_3: p_j = \phi_2,$$
where $\phi_1$ denotes the DLT probability that is deemed substantially lower than the target (i.e., underdosing) such that dose escalation should be made, and $\phi_2$ denotes the DLT probability that is deemed substantially higher than the target (i.e., overdosing) such that dose de-escalation is required. Thus, the correct decision under $H_1$, $H_2$, and $H_3$ is stay, escalation, and de-escalation, respectively; and other decisions are incorrect. For example, under $H_1$,  escalation or de-escalation are incorrect decisions. Liu and Yuan (2015) showed that optimal dose escalation and de-escalation boundaries that minimize incorrect decisions are given by
\begin{eqnarray}
\lambda_{e} &= & {\rm max}\left\{0, \quad  \frac{\mbox{log}\left(\frac{1-\phi_1}{1-\phi}\right)+n_j^{-1}\mbox{log}\left(\frac{\pi_{2j}}{\pi_{1j}}\right)}{\mbox{log}\left\{\frac{\phi(1-\phi_1)}{\phi_1(1-\phi)}\right\}}\right\}, \notag \\
 \lambda_{d} &=& {\rm min}\left\{1, \quad  \frac{\mbox{log}\left(\frac{1-\phi}{1-\phi_2}\right)+n_j^{-1}\mbox{log}\left(\frac{\pi_{1j}}{\pi_{3j}}\right)}{\mbox{log}\left\{\frac{\phi_2(1-\phi)}{\phi(1-\phi_2)}\right\}} \right\} ,\label{BOIN0}
\end{eqnarray}
where $\pi_{kj}=\mbox{Pr}(H_{k})$ is the prior probability that the hypothesis $H_{k}$ is true at dose level $j$, where $k=1,2,3$. As a result, BOIN is the optimal design with the lowest decision error rate among all nonparametric designs. Liu and Yuan (2015) recommended default values $\phi_1=0.6\phi$ and $\phi_2=1.4\phi$, which lead to desirable operating characteristics and the decision rule that fits most clinical practices.

When there is no reliable prior information available, we can take the non-informative prior approach and assign the equal probability to each of the three hypotheses being true, i.e., $\pi_{1j}=\pi_{2j}=\pi_{3j}=1/3$. Then, the optimal boundaries (\ref{BOIN0}) become
\begin{equation}
\lambda_e^*=\frac{\mbox{log}\left(\frac{1-\phi_1}{1-\phi}\right)}{\mbox{log}\left\{\frac{\phi(1-\phi_1)}{\phi_1(1-\phi)}\right\}} \hspace{1cm} \mbox{and}\hspace{1cm} \lambda_d^*=\frac{\mbox{log}\left(\frac{1-\phi}{1-\phi_2}\right)}{\mbox{log}\left\{\frac{\phi_2(1-\phi)}{\phi(1-\phi_2)}\right\}},
\end{equation}
which have the desirable feature that they are independent to the dose level $j$ and the number of patients treated $n_j$. This means that the same pair of dose escalation and de-escalation boundaries $(\lambda_d, \lambda_e)$ can be used throughout the trial to make the decision of dose escalation and de-escalation, making the BOIN design particularly simple to implement. That is, if $\hat{p}_j < \lambda_e^*$, escalate the dose; if $\hat{p}_j > \lambda_d^*$, de-escalate the dose; otherwise, stay at the current dose.

When prior information is available, we propose the following procedure to incorporate it into the design:

\begin{enumerate}
	\item Elicit skeleton $(q_1, \cdots, q_J)$ and corresponding PESS $(n_{01},\cdots, n_{0J})$, where $n_{0j}$ is the desirable PESS for dose level $j$, $j=1, \cdots, J$.
	\item Determine the informative prior for $H_k$, i.e., $\pi_{kj}$, as
	\begin{align}
	\pi_{kj} = \sum_{x=0}^{n_0}\frac{\phi_{k}^{x}(1-\phi_{k})^{n_0-x}}{\sum_{k'=1}^{3}\phi_{k'}^{x}(1-\phi_{k'})^{n_0-x}}\binom{n_0}{x} q_j^{x} (1-q_j)^{n_0-x} \label{pi}.
	\end{align}

	\item Make the decision of dose escalation or de-escalation according to the boundaries given in (\ref{BOIN0}) with $\pi_{kj}$ determined in step 2.
\end{enumerate}
The derivation of $\pi_{kj}$ in step 2 is provided in Section \ref{sec:derivation}. We refer to the resulting design (with informative prior) as iBOIN. 

\begin{figure}[ht!]
	\begin{center}
		\includegraphics[scale=.65]{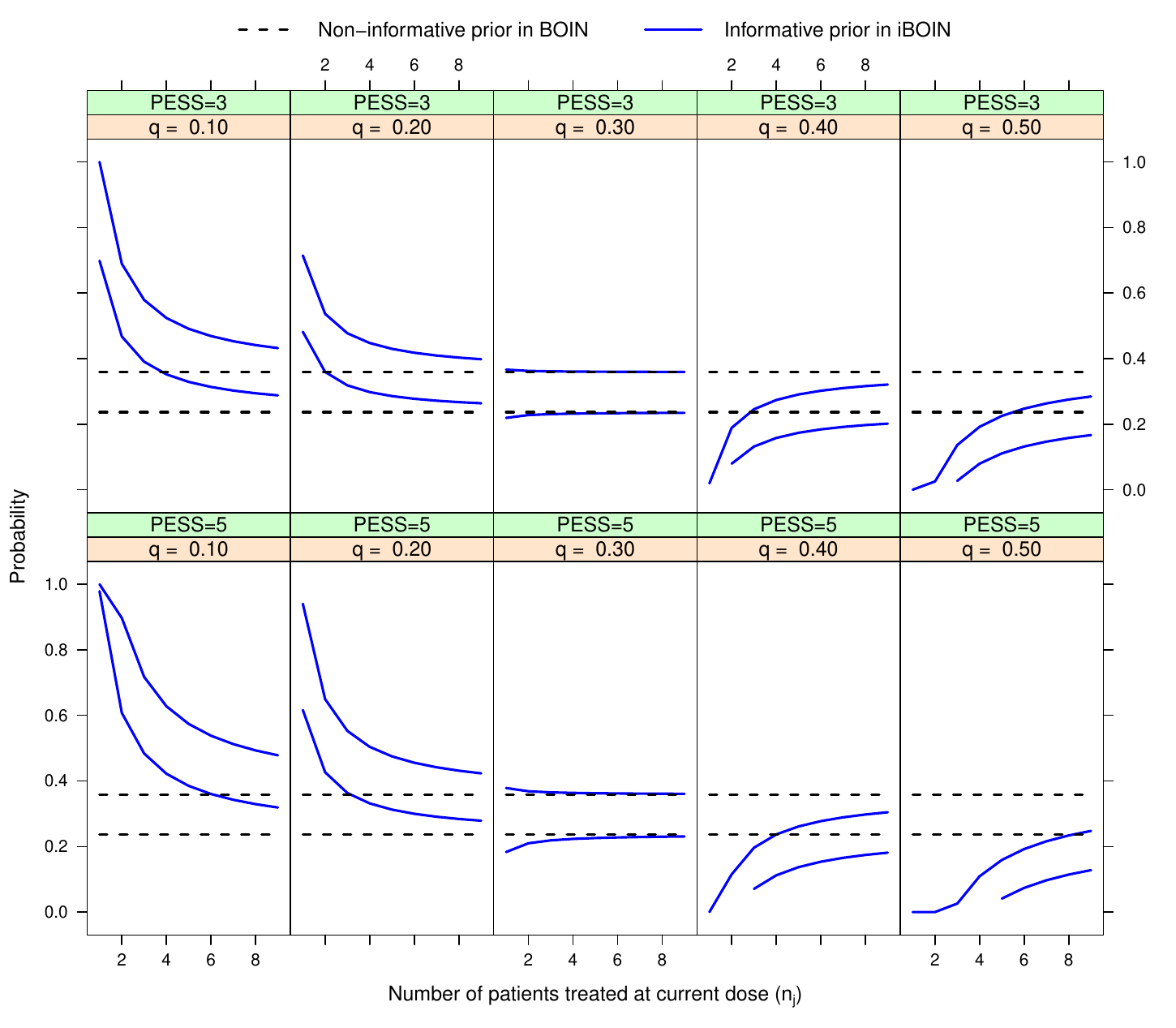}
		\caption{Escalation and de-escalation boundaries $(\lambda_{e},\lambda_{d})$ of iBOIN given different prior DLT probability ($q$) and PESS = 3 or 5, in comparison to the boundaries determined using a non-informative prior in standard BOIN.}
		\label{fig:bounds}
	\end{center}
\end{figure}

Because of the incorporation of the informative prior information, the escalation and de-escalation boundaries $\lambda_e$ and $\lambda_d$ of iBOIN depend on the dose level $j$, as well as $n_j$. Figure \ref{fig:bounds} contrasts the boundaries under a non-informative prior and those under an informative prior for a trial with 5 doses and an elicited skeleton (0.10, 0.19, 0.30, 0.42, 0.54), when the target DLT probability is 0.3 and PESS is 3 or 5. For example, because the prior information says that the lowest dose is below the true MTD (with the prior DLT probability of 0.10), its escalation boundary $\lambda_e$ is higher than that of the non-informative prior to encourage dose escalation. On the contrary, because the prior information says the highest dose is above the MTD (with the prior DLT probability of 0.54), its de-escalation boundary $\lambda_d$ is lower than that of the non-informative prior to encourage dose de-escalation. The informative decision boundaries approach to those in stadnard BOIN (with noninformative prior),  when the number of patients treated increases (i.e., data start to override prior information). iBOIN becomes the standard BOIN, when $(n_{01},\cdots, n_{0J})=0$. 

Compared to CRM, iBOIN is more flexible and allows users to accurately incorporate prior information by specifying a PESS for each dose. For example, given a phase I trial with 5 doses, if historical data provide more information on the first 2 doses than the last 2 doses and most information on dose level 3, we could specify the 5 doses' PESS as (3, 3, 6, 1, 1) to reflect that. As described previously, this is extremely difficult, if not impossible, under CRM. 

The other advantage of iBOIN is that the dose escalation and de-escalation rule can be pre-tabulated and included in the trial protocol.  Table \ref{tab:boiny} shows the decision table of iBOIN with skeleton (0.10, 0.19, 0.30, 0.42, 0.54) and the effective sample size $n_{01}=\cdots=n_{05}=3$. This decision table is equivalent to the rule based on $\lambda_e$ and $\lambda_d$, but easier to use in practice. Users need only identify the row corresponding to the current dose level, and then they can use the boundaries listed in that row to easily make the decision of dose escalation and de-escalation. In summary, the iBOIN design can be described as follows:
\begin{enumerate}
	\item Patients in the first cohort are treated at
	the lowest dose $d_1$, or the physician-specified dose.
	\item	Given data $(n_j, y_j)$ observed at the current dose level $j$, make the decision of escalation/de-escalation according to the iBOIN decision table (e.g., Table \ref{tab:boiny}) for treating the next cohort of patients.
	\item 	Repeat step 2 until the prespecified maximum sample size is reached, and then select the MTD as the dose whose isotonically transformed estimate of ${p}_j$ is closest to $\phi$.
\end{enumerate}

For the purpose of overdose control, following BOIN, the iBOIN design imposes a dose elimination rule:  if $\mbox{Pr}(p_j > \phi \,|\, y_j, n_j)>0.95$ and $n_j\ge 3$, dose level $j$ and higher are eliminated from the trial, where ${\rm Pr}(p_j > \phi ~|~ n_j, y_j)$ is evaluated based on the beta-binomial model with the $\mbox{uniform}(0,1)$ prior. As the objective of the dose elimination rule is to protect patients from excessively toxic doses, it is sensible to use the uniform prior to evaluate this rule to avoid potential bias due to potential misspecification of the prior. The trial is terminated if the lowest dose level is eliminated.

At the end of the trial, iBOIN uses the isotonic estimate of $p_j$ to select the MTD (i.e., step 3). As determining dose escalation/de-escalation and selecting the MTD are two independent components, when the trial is completed, other methods can also be used to determine the MTD. For example, when desirable, we can fit a dose-toxicity model (e.g., a logistic model) as CRM to select the MTD.

\subsection{Incorporate prior information in keyboard/mTPI-2 design}
The keyboard design is another model-assisted design, which was developed to address the overdosing issue of the mTPI design.  
Guo et al. \cite{guo2017bayesian} proposed a modification of mTPI, known as mTPI-2, which is statistically equivalent to the keyboard design, but less transparent and relying upon a perplexing statistical concept and method (e.g., Occam’s razor and model selection). Thus, we only present the keyboard design. The methodology described below is directly applicable to mTPI and mTPI-2.

The keyboard design starts by specifying a proper dosing interval ${\cal I}^*=(\delta_1, \delta_2)$, referred to as the ``target key," and then populates this interval toward both sides of the target key, forming a series of keys of equal width that span the range of 0 to 1. For example, given a target rate of $\phi = 0.30$, the proper dosing interval or target key may be defined as (0.25, 0.35), then on its left side, we form 2 keys of width 0.1, i.e., (0.15, 0.25) and (0.05, 0.15); and on its right side, we form 6 keys of width 0.1, i.e., (0.35, 0.45), (0.45, 0.55), (0.55, 0.65), (0.65, 0.75), (0.75, 0.85) and (0.85, 0.95). We denote the resulting intervals/keys as ${\cal I}_1, \cdots, {\cal I}_K$.

The keyboard design assumes a beta-binomial model,
\begin{eqnarray}
y_j \,|\, n_j, p_j & \sim & \mathrm{Binom}(n_j,p_j) \notag \\
p_j & \sim & \mathrm{Beta}(a_j, b_j), \label{betabinomial}
\end{eqnarray}
where $a_j$ and $b_j$ are hyperparameters.
The posterior distribution of $p_j$ arises as
\begin{equation}
p_j \,|\, D_j \sim \mathrm{Beta}(y_j+a_j, n_j-y_j+b_j), \; \mathrm{for} \; j=1,\ldots,J. \label{bbpost}
\end{equation}
By default, the keyboard design set $a_j=b_j=1$ to obtain a uniform prior.
To make the decision of dose escalation and de-escalation, given the observed data $D_j=(n_j, y_j)$ at the current dose level $j$, the keyboard design identifies the interval ${\cal I}_{\rm max}$ that has the largest posterior probability, i.e.,
$${\cal I}_{\max}  = {\rm argmax}_{{\cal I}_1, \cdots, {\cal I_K}} \{ {\rm Pr}( p_j \in {\cal I}_k \,|\, D_j); \,\, k=1, \cdots, K\}.$$
${\cal I}_{\rm max}$ represents the interval within which the true value of $p_j$ is most likely located, referred to as the ``strongest" key by Yan et al. \cite{yan2017keyboard}. Suppose $j$ is the current dose level. The keyboard design determines the next dose as follows.
\begin{itemize}
	\item Escalate the dose to level $j+1$, if the strongest key is on the left side of the target key. 
	\item Stay at the current dose level $j$, if the strongest key is the target key.
	\item De-escalate the dose to level $j-1$, if the strongest key is on the right side of the target key.
\end{itemize}
The trial continues until the prespecified sample size is exhausted, and the MTD is selected based on isotonic estimates of $p_j$. During the trial conduct, the keyboard design imposes the same dose elimination/early stopping rule as the BOIN design.

As in the beta-binomial model (\ref{betabinomial}), $a_j+b_j$  can be interpreted as the PESS. We propose the following procedure to incorporate prior information into the keyboard design:
\begin{enumerate}
	\item Elicit skeleton $(q_1, \cdots, q_J)$ and corresponding PESS $(n_{01},\cdots, n_{0J})$, where $n_{0j}$ is the desirable PESS for dose level $j$, $j=1, \cdots, J$.

	\item Determine hyperparameter $a_j$ and $b_j$ in the beta prior (\ref{betabinomial}) as follows:
	\begin{align}
	a_j = n_{0j} q_j; \qquad b_j = n_{0j} (1-q_j),  \qquad j=1, \cdots,J
	\end{align}
	\item Make dose escalation and de-escalation based on the resulting posterior given by equation (\ref{bbpost}).
\end{enumerate}
We refer to the keyboard design with an informative prior as the iKeyboard design. Given a fixed maximum sample size, all possible outcomes $D_j=(n_j, y_j)$ can be enumerated, and for each possible outcome, the posterior distribution $f(p_j|D_j)$ can be calculated. Therefore, the dose escalation/de-escalation rule of iKeyboard can be tabulated. The decision table for the iKeyboard design can also be pre-tabulated, similar to iBOIN. The above approach is directly applicable to the mTPI design for incorporating prior information. As the keyboard/mTPI-2 design outperforms the mTPI design in both safety and accuracy (Yan et al., \cite{yan2017keyboard}; Guo et al.\cite{guo2017bayesian}; Zhou et al. \cite{zhou2018comparative}), we will not discuss mTPI.

\setlength{\tabcolsep}{12pt}
\renewcommand{\arraystretch}{1}

\subsection{Robust prior}\label{sec: robust_prior}
The performance of aforementioned designs is affected by whether the informative prior is correctly specified. When the informative prior is correctly specified, it improves the accuracy of identifying the MTD. However, when the informative prior is seriously misspecified, 
it may compromise the accuracy of identifying the MTD. 
In the numerical study described later, we found that the impact of the misspecification depends on both the location of the prior MTD and the location of the true MTD.
For example, consider two cases: in case 1, the prior sets dose level 3 as the MTD, while the true MTD is dose level 5; and in case 2, the prior sets dose level 1 as the MTD, while the true MTD is dose level 3. Although both priors are misspecified by 2 dose levels, iBOIN and iKeyboard designs have a lower probability of identifying the true MTD in case 1 than in case 2. This is because in case 2, the prior MTD is dose level 1; there is sufficient sample size to override the prior and escalate to find the true MTD. In contrast, in case 1, because the MTD is dose level 5, the sample size is often exhausted before enough data are accumulated at dose level 3 (i.e., the prior MTD) to override the prior, thus fails to find the true MTD with a high probability.  

This observation motivated us to propose a robust prior, which is useful when there is a great amount of uncertainty regarding the prior information. Given the elicited skeleton $(q_1, \cdots, q_J)$ with dose level $j^*$ as the prior estimate of the MTD (i.e., $q_{j^*}=\phi$), the robust prior is the same as the prior described above when $j^*< J/2$, but modify PESS to $(n_{01},\cdots, n_{0j^*}, 0, \cdots, 0)$ when $j^*\ge J/2$. In other words, when prior MTD $j^*\ge J/2$, the robust prior uses informative prior information for the dose up to the prior estimate of the MTD, and after that it uses the non-informative prior. This modification facilitates overriding the prior when the data conflict with the prior,  and thus alleviates the impact of prior misspecification. Our simulation study described later shows that the iBOIN and iKeyboard designs are robust to moderate misspecification of priors, and using the robust prior proves their robustness when the prior is severely misspecified. 

Another way to robustify the prior is to use mixture \cite{schmidli2014robust}. Let $\pi_{inf}$ and $\pi_{non}$ generically denote the informative prior (obtained based on the skeleton and PESS) and the non-informative prior described previously. The mixture prior is given by $\pi_{mix} = w \pi_{inf} + (1-w) \pi_{non}$, where $0\le w \le 1$ is a prespecified mixture proportion. If the prior information is reliable, we assign $w$ a large value (e.g., $w=0.9$); and if the prior information has high uncertainty, we assign $w$ a small value (e.g., $w=0.5$). Numerical study shows that the mixture prior does not perform as well as the aforementioned robust prior (see Appendix \ref{sec:mixture} for simulation results), thus hereafter we focus on the latter.

\subsection{Choose PESS }\label{sec:choosePESS}

PESS should be chosen to reflect the appropriate amount of prior information to be incorporated, which depends on the reliability of the prior information and varies from trial to trial. When there is strong evidence that the prior is most likely correctly specified, it is appropriate to use a large PESS to borrow more information; when there is a great amount of uncertainty regarding whether the prior is most likely correctly specified, we may use a small PESS to avoid bias. In practice, there is often sizable uncertainty on the reliability of the prior information. Thus, PESS should be chosen carefully to achieve an appropriate balance between design performance and robustness. Using a large PESS improves the design performance (i.e., the accuracy to identify the MTD) when the prior is correctly specified, but may lead to a substantial loss of performance when the prior is severely misspecified. Based on numerical study, we recommend $\mbox{PESS} \in [1/3(N/J),  1/2(N/J)]$ as the default value that improves trial performance while maintaining reasonably robust. For example, when $J=5$ and $N=30$, the recommended value for PESS is $n_{0j}=2$ or 3 (i.e., across 5 doses, the total PESS is 10 or 15). The value of $n_{0j}$ can be further calibrated by simulation using the software described in the next section.

\section{Software}\label{sec:app}
We have developed the online software ``BOIN Suite" to allow users to design trials, conduct simulations, and generate protocol templates. The software has an intuitive graphical user interface and rich documents to help with navigating through the process, see Figure \ref{fig:app} for the user interface of the software, which is freely available at {\it https://www.trialdesign.org}. A trial can be easily designed via the following three steps. 
\begin{description}
 \item[Step 1.] Specify the design parameters, e.g., sample size, cohort size, target DLT probability, skeleton, and PESS.
\item[Step 2.] Use the software to produce a decision table and design diagram, and conduct simulation to obtain the operating characteristics of the design. The software also generates sample texts and protocol templates to facilitate writing the protocol.
\item[Step 3.] Use the design decision table to conduct the trial. 
\end{description}
After a trial completes, use the app to select the MTD. 

\section{Simulation}\label{sec:simulation}
We conducted extensive simulations to evaluate the operating characteristics of the proposed designs. We assumed $J=5$ doses and the target DLT probability $\phi=0.3$. The maximum sample size was $N=30$ with a cohort size of 3. We considered the CRM with an informative prior (denoted as iCRM), iBOIN, and iKeyboard, as well as their counterparts with a non-informative prior. We also considered BOIN and Keyboard designs with the robust prior, denoted as iBOIN$_R$ and iKeyboard$_R$, respectively. For iBOIN and iKeyboard, we set PESS $n_{0j}=3$ for $j=1, \cdots 5$; and for iCRM, the prior is chosen such that the PESS at the prior MTD is 3. All the designs use the same skeletons (i.e., the prior DLT probabilities), which are provided in Table \ref{scenarios}.

\subsection{Fixed scenarios}
We evaluated the performance of the designs in ten scenarios, as shown in Table \ref{scenarios}. In the first five scenarios, the MTD was located at dose level 1, 2, 3, 4 and 5, respectively; and the prior MTD was correctly specified and matched the true MTD. To reflect the practice, we did not assume that the prior (at each dose level) exactly matched the truth. Here, we called a prior correctly specified if the prior MTD matched the true MTD. Scenarios 6-10 considered the cases that the prior was misspecified. Specifically, in scenarios 6 and 7,  the prior MTD was one level off from the true MTD, and in scenarios 8-10,  the prior MTD was two levels off from the true MTD. 

Table \ref{OC_target30_summary} shows the results, including (1) percentage of correct selection (PCS), defined as the percentage of simulated trials in which the MTD is correctly identified; (2) percentage of patients treated at MTD;  (3) percentage of patients treated above MTD; (4) risk of overdosing, defined as the percentage of simulated trials that assigned 50\% or more patients to doses above MTD; and (5) risk of poor allocation, defined as the percentage of simulated trials that assigned less than six patients to MTD. As noted by Zhou et al. \cite{zhou2018accuracy}, metrics (4) to (5) measure the reliability of the design, i.e., the likelihood of a design demonstrating extreme problematic behaviors (e.g., treating 50\% or more patients at toxic doses, or fewer than six patients at the MTD), which are of great practical importance.  Note that the percentage of patients overdosed (i.e., metric (3)) does not cover the risk of overdosing (i.e., metric (4)). Two designs can have a similar percentage of patients overdosed, but rather different risks of overdosing 50\% of the patients.

In scenarios 1 to 5,  the prior was correctly specified. iCRM and iBOIN outperformed their counterparts that use non-informative priors. Specifically, compared to CRM, iCRM improved PCS and the percentage of patients treated at MTD by 2-8\% and 3-6\%, respectively. Compared to BOIN, iBOIN improved PCS and the percentage of patients treated at MTD by 5-8\% and 5-7\%, respectively. iCRM and iBOIN yielded comparable PCS and the percentage of patients assigned to MTD, but iBOIN was more reliable with a lower risk of overdoing. For example, in scenarios 2 and 3, the risk of overdosing for iBOIN was about half and one fourth of that for iCRM, respectively. Compared to its non-informative counterpart, iKeyboard had a 5-13\% increase in PCS, but the percentage of patients treated at MTD was often lower and the risk of overdosing was substantially increased by more than 10\% in most scenarios. iBOIN$_R$ and iKeyboard$_R$ yielded similar performances to iBOIN and iKeyboard, respectively. 

Scenarios 6 and 7 considered the cases where the prior was misspecified, with the prior MTD being one level off from the true MTD. iCRM and iBOIN were robust to this moderate prior misspecification and outperformed their counterparts. For example, in scenario 6, the PCS of iCRM and iBOIN were 57.3\% and 58.6\%, respectively; this was 6.6\% and 7.1\% higher than CRM and BOIN. Compared to iCRM, iBOIN had a lower overdose risk. Scenarios 8 and 9 examined the cases where the prior was severely misspecified, with the prior MTD being two levels off from the true MTD. When the prior MTD was higher than the true MTD (i.e., scenario 8), iCRM and iBOIN performed well, yielding performances comparable to their non-informative counterparts.  The PCS of iKeyboard was lower than keyboard. When the prior MTD was lower than the true MTD (i.e., scenario 9), the prior misspecification had more impact on the performance of the designs. The PCS of iCRM and iBOIN was lower than their non-informative counterparts. Scenario 9 was a difficult scenario, because the true MTD was the highest dose. The sample size was often exhausted before enough data were accumulated to overcome the misspecified prior to reach the highest dose (i.e., MTD). In this scenario, iCRM performed better than iBOIN because the iCRM tended to escalate the dose more aggressively, as demonstrated by its relatively high risk of overdosing. The proposed robust prior addressed this issue. iBOIN$_R$ yielded a higher PCS, comparable to standard BOIN. In the case that the prior MTD was two levels lower than the true MTD, but the true MTD was not the highest dose, iCRM and iBOIN outperformed their non-informative counterparts (see scenario 10).

\subsection{Random scenarios}
To validate the above results, we repeated the simulation using a large number of scenarios randomly generated using a pseudo-uniform algorithm \cite{clertant2017semiparametric}. Two large sets of random scenarios were constructed. The first set was used to examine the operating characteristics of the designs when the prior was correctly specified. We generated 2000 random scenarios with MTD located at dose level 1, 2, 3, 4, and 5, with equal probability, and we assumed that the prior MTD was correctly specified for each of the scenarios. The second set was used to evaluate the performance of the designs when the prior was misspecified. We considered two types of misspecification: the prior MTD was one level off from the true MTD, and the prior MTD was two levels off from the true MTD. For each type of misspecification, we generated 4000 random scenarios with half of them having the prior MTD (one or two levels) lower than the true MTD, and the other half having the prior MTD (one or two levels) higher than the true MTD.  We simulated 2000 trials for each scenario. Details on random scenario generation and configuration are provided in Section \ref{sec:generate_random}.

\begin{figure}[ht]
	\includegraphics[scale=0.7]{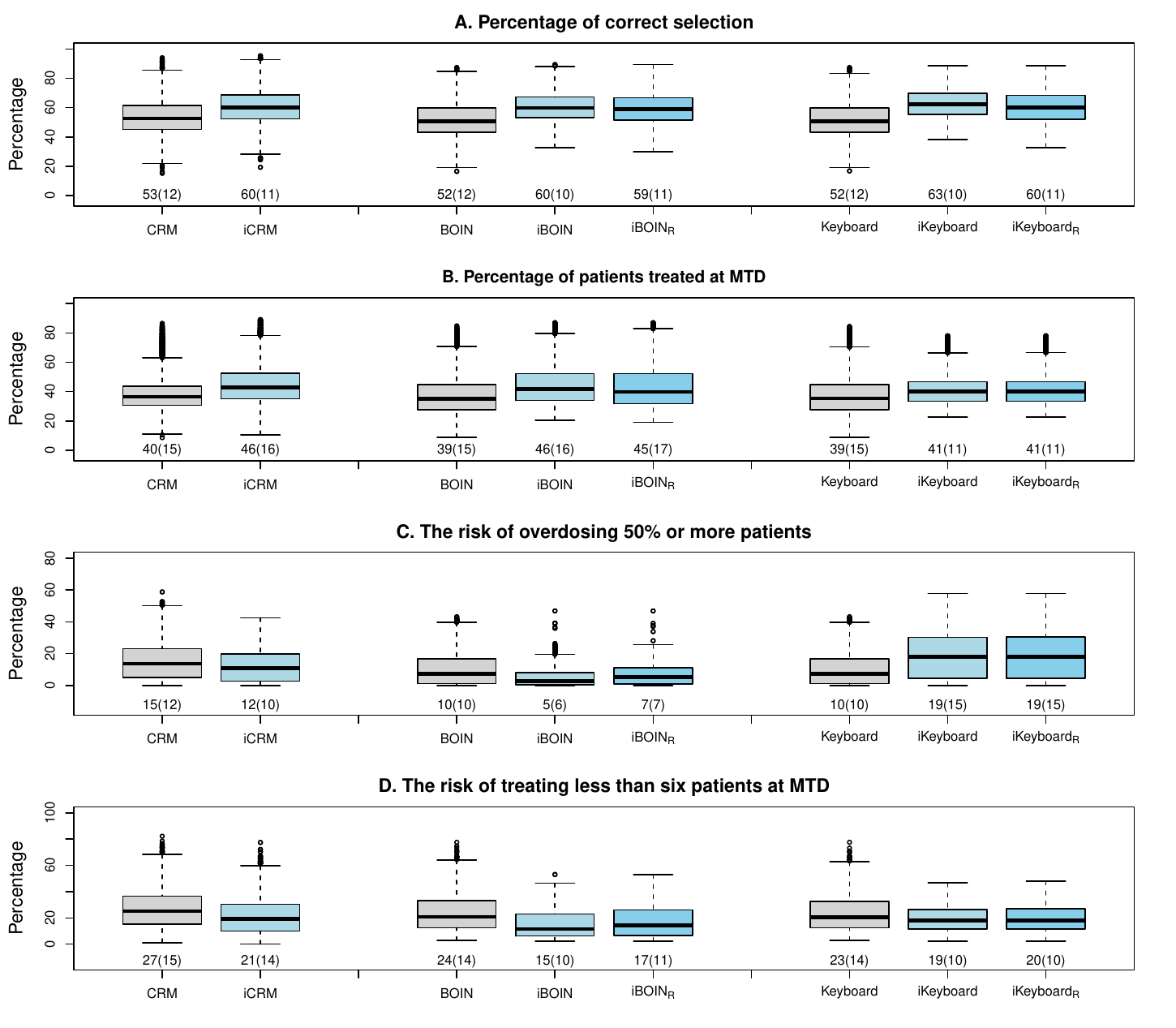}
	\caption{Operating characteristics of iCRM,  iBOIN, and iKeyboard, in comparison to their counterparts with non-informative priors, under 2000 random scenarios when the prior is \textbf{correctly specified}. $\mbox{iBOIN}_R$ and $\mbox{iKeyboard}_R$ are iBOIN and iKeyboard using robust priors, respectively. The number under each boxplot is the average value with the standard deviation shown in parenthesis.}
	\label{fig:OC30boxplot}
\end{figure}

Figure \ref{fig:OC30boxplot} shows the simulation results when the prior MTD was correctly specified. The findings were generally consistent with the results based on the fixed scenarios. That is, iCRM and iBOIN outperformed their non-informative counterparts with a higher PCS and a higher percentage of patients assigned to MTD. For example, averaging over 2000 random scenarios, the PCS of iCRM was 7\% higher than that of the CRM, and the PCS of iBOIN is 8\% higher than that of BOIN. iCRM and iBOIN yielded similar PCS and percentage of patients to the MTD, but iBOIN had a lower risk of overdosing and poor allocation. The iKeyboard design yielded a higher PCS than its non-informative counterpart, but increased the risk of overdosing due to its aggressive dose escalation. 

\begin{figure}[ht]
	\includegraphics[scale=0.7]{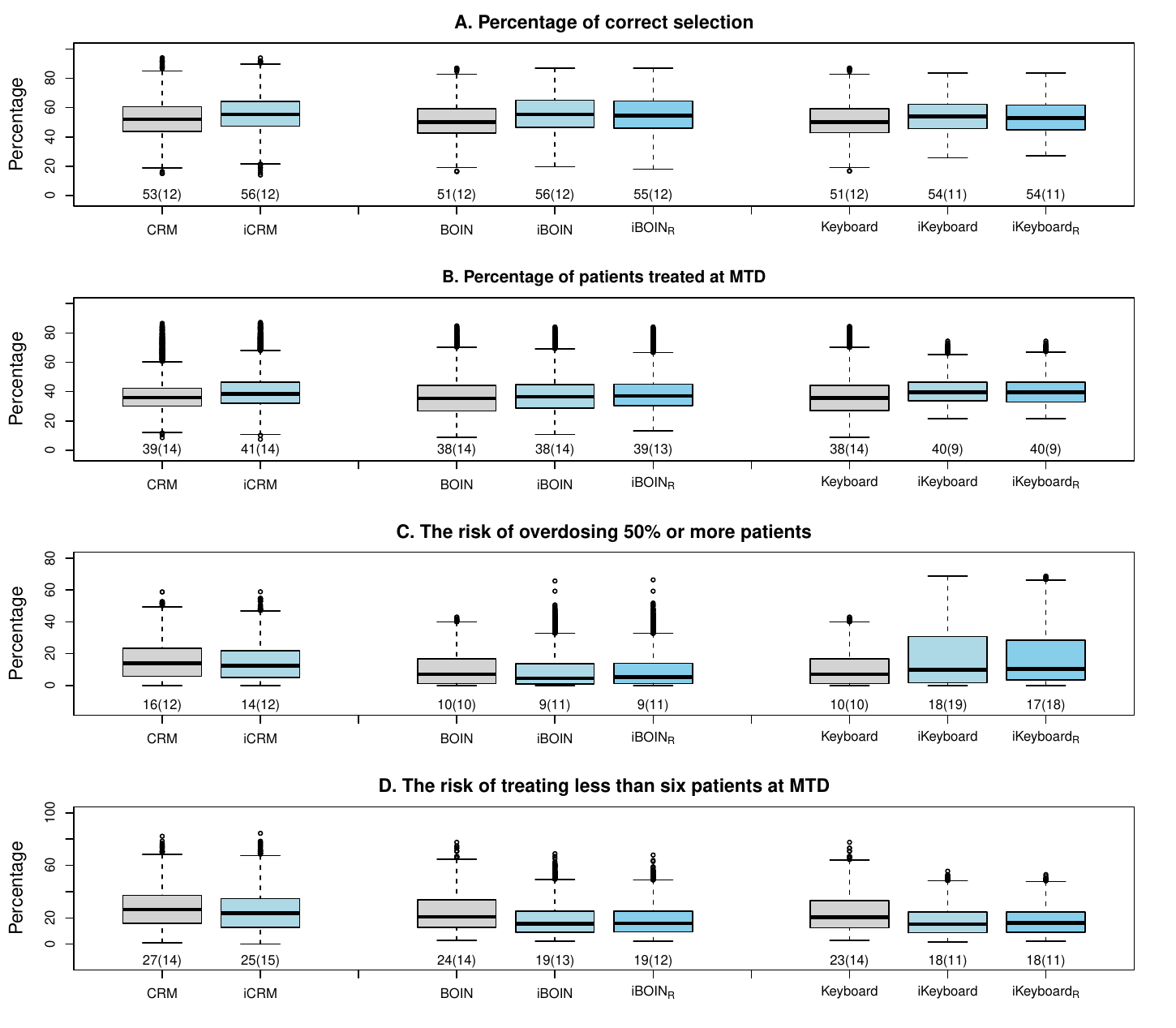}
	\caption{
	Operating characteristics of iCRM,  iBOIN and iKeyboard, in comparison to their counterparts with non-informative priors, under 4000 random scenarios when the prior MTD is \textbf{one dose off} from the true MTD. $\mbox{iBOIN}_R$ and $\mbox{iKeyboard}_R$ are iBOIN and iKeyboard using robust priors, respectively. The number under each boxplot is the average value with the standard deviation shown in parenthesis.	}
	\label{fig:OC30boxplot_oneoff}
\end{figure}

\begin{figure}[ht!]
	\includegraphics[scale=0.7]{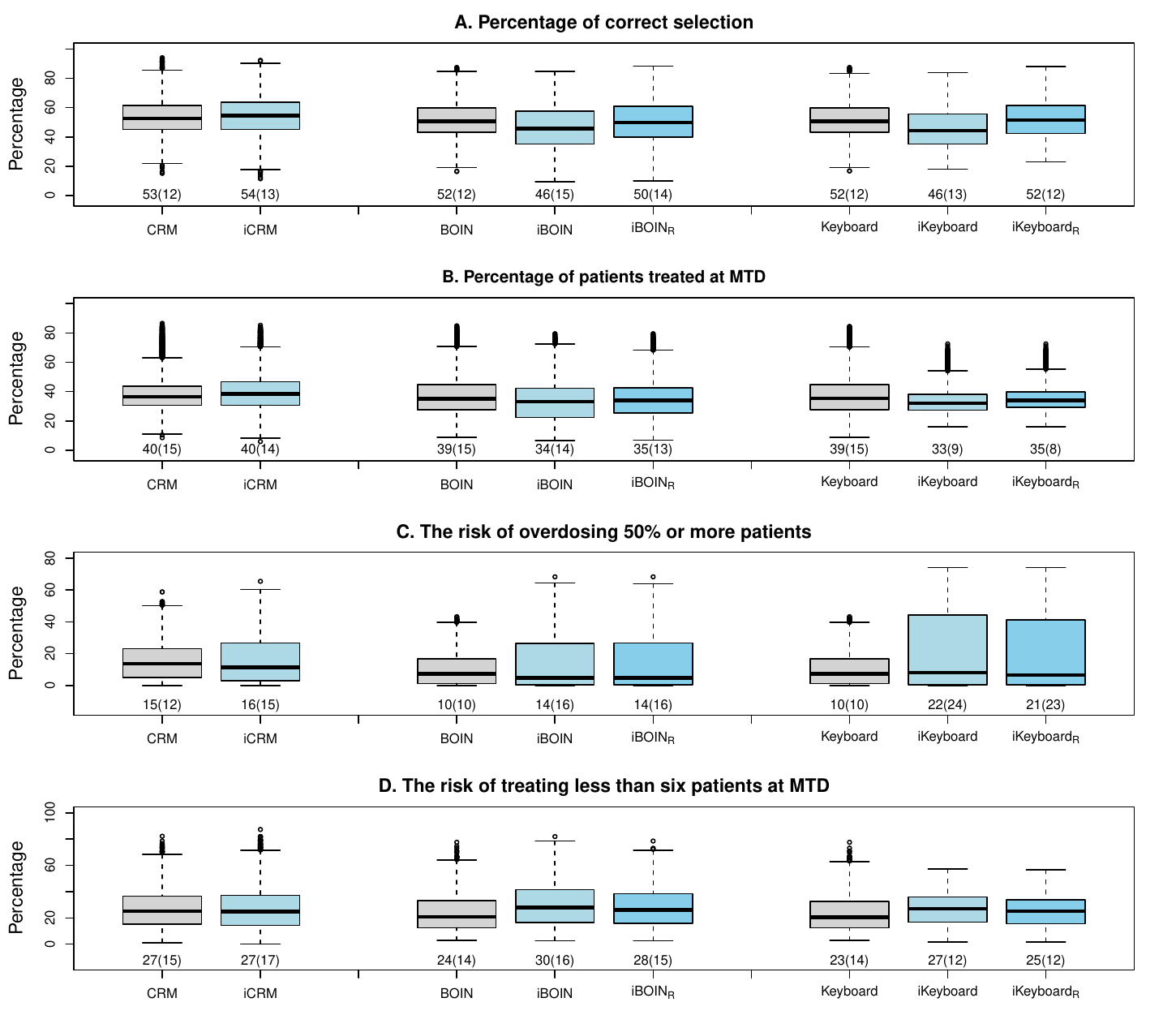}
	\caption{Operating characteristics of  iCRM,  iBOIN and iKeyboard, in comparison to their counterparts with non-informative priors, under 4000 random scenarios when the prior MTD is \textbf{two doses off} from the true MTD. $\mbox{iBOIN}_R$ and $\mbox{iKeyboard}_R$ are iBOIN and iKeyboard using robust priors, respectively. The number under each boxplot is the average value with the standard deviation shown in parenthesis.}
	\label{fig:OC30boxplot_twooff}
\end{figure}

Figure \ref{fig:OC30boxplot_oneoff} shows the simulation results when the prior was misspecified by one dose level. iCRM and iBOIN proved robust to such moderate prior misspecification. The PCS of iCRM and iBOIN were both 56\%, offering 3\% and 5\% improvement over their non-informative counterparts, respectively. The risk of overdosing and poor allocation for iCRM were respectively 5\% and 6\% higher than that for iBOIN. The iKeyboard design offered 3\% improvement over its non-informative counterpart, but the risk of overdosing was 8\% higher. When the prior was severely misspecified with the prior MTD being two levels off from the true MTD (see Figure \ref{fig:OC30boxplot_twooff}), iCRM was more robust than iBOIN; however, by using the proposed robust prior, iBOIN$_R$ showed competitive performance. In practice, when the prior is likely to be severely misspecified, using prior information should be avoided in favor of the more sensible non-informative prior.

\subsection{Unequal prior information across doses}
\begin{figure}
	\includegraphics[scale=0.7]{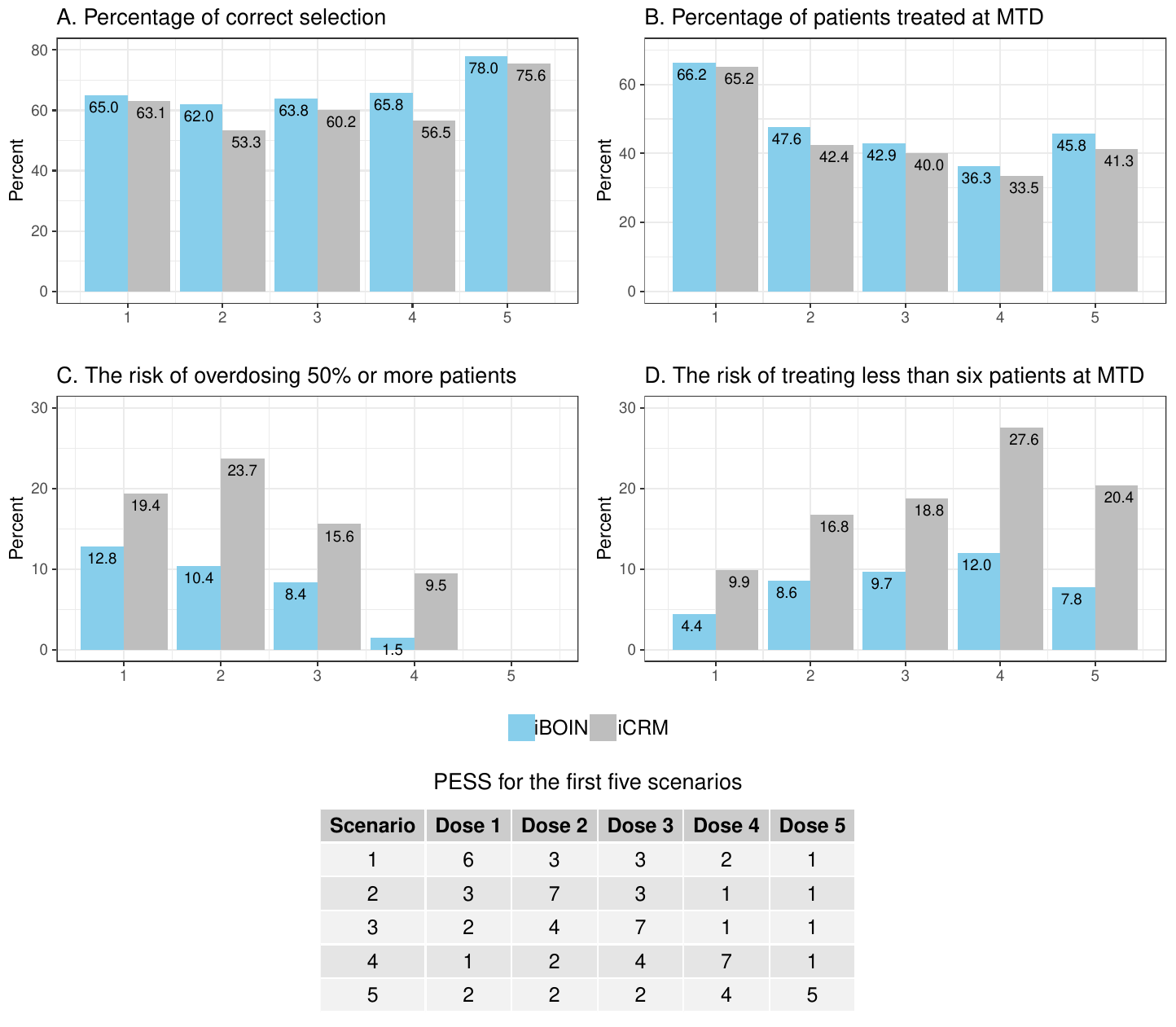}
	\caption{Operating characteristics of iBOIN and iCRM when different amount of prior information (i.e., PESS) is available for different doses under scenarios 1 to 5.}
	\label{fig:iCRM_iBOINp}
\end{figure}

Lastly, we briefly investigated the case that different amounts of prior information were available for different doses. We assumed that more prior data were available at lower doses than higher doses, and more prior data were available around the prior MTD, as we often observed in (historical) phase I trials. As described in Section \ref{sec:method}, iBOIN can easily accommodate this by specifying different PESS at different doses. Figure \ref{fig:iCRM_iBOINp} shows the simulation results under scenarios 1 to 5. We controlled the total PESS over five doses as the same for iBOIN and iCRM (i.e., CRM). We see that, compared to CRM, iBOIN offered a higher PCS and allocated a larger percentage of patients at the MTD, as well as a lower risk of overdosing and poor allocation. CRM does not allow for specifying dose-specific PESS as it uses a single parameter to control prior information in all doses, thus it cannot take full advantage of the prior information.

\section{Conclusion}\label{sec:discussion}

In this paper, we propose a unified framework to incorporate historical data or real-world evidence to improve the efficiency of phase I trial designs, especially model-assisted designs. By using skeleton and PESS, our method is intuitive and easy to interpret. More importantly, our approach maintains the hallmark of model-assisted designs: simplicity---the dose escalation/de-escalation rule can be tabulated prior to the trial conduct. For example, implementing the proposed iBOIN only involves a simple comparison of the number of DLTs observed at the current dose with the prespecified dose escalation and de-escalation boundaries (e.g., Table \ref{tab:boiny}).  Extensive simulation studies show that the proposed method, in particular iBOIN, can effectively incorporate prior information and yield comparable performance as the model-based CRM design, but with greater reliability. Moreover, iBOIN is more transparent and easier to implement. In addition, iBOIN has greater flexibility and allows for specifying dose-specific prior information to more accurately reflect available prior information. The iBOIN design is generally robust to prior misspecification. When there is a high likelihood that the prior is severely misspecified, the proposed robust prior can be used with iBOIN to enhance its robustness. Actually, in this case, there is little rationale to incorporate prior information and it is more appropriate to use a non-informative prior. When non-informative prior is used, iBOIN becomes standard BOIN. Freely available software is provided at {\it https://www.trialdesign.org} to facilitate the use of proposed designs.

The proposed methodology requires prespecification of skeleton and PESS. Investigators often have good knowledge on the skeleton, e.g., obtained by fitting a model to historical data, but less knowledge on the PESS. Thus in some applications, investigators may have difficulty to specify the PESS or worry about the accuracy of the PESS. In general, we do not regard this as an issue in practice. Our numerical study shows that the proposed designs are remarkably robust. In addition, as it is undesirable to let prior information dominate the trial data, the range for the reasonable PESS actually is narrow, typically within [0, 6], given small sample size of phase I trials. In the case that investigators have difficulty to choose the PESS,  our recommended PESS $\in [1/3(N/J ), 1/2(N/J )]$ is a good choice, in particularly used with the proposed robust prior. Other approaches might be taken to further alleviate this issue. For example, rather than eliciting a value of PESS, we can ask clinicians to provide a prior distribution of the PESS, e.g., Pr(PESS=1)=0.2, Pr(PESS=2)=0.6 and Pr(PESS=3)=0.2, to incorporate their uncertainty on the PESS. The other possible approach is to adjust the PESS adaptively using two-stage design. The first stage uses the prespecified PESS. At the end of the first stage, if the observed interim data show the evidence of conflicting with the skeleton, we may discount the PESS for conducting the second stage of dose-finding. These approaches warrant further research. 

 \clearpage

\bibliographystyle{unsrt}
\bibliography{references}

\begin{thebibliography}{10}

\bibitem{us2019submitting}
US~Food, Drug Administration, et~al.
\newblock Submitting documents using real-world data and real-world evidence to
  fda for drugs and biologics: Guidance for industry: Draft guidance.
\newblock {\em Rockville, MD: US Food and Drug Administration}, 2019.

\bibitem{zohar2011approach}
Sarah Zohar, Sandrine Katsahian, and John O'Quigley.
\newblock An approach to meta-analysis of dose-finding studies.
\newblock {\em Statistics in Medicine}, 30(17):2109--2116, 2011.

\bibitem{liu2015bridging}
Suyu Liu, Haitao Pan, Jielai Xia, Qin Huang, and Ying Yuan.
\newblock Bridging continual reassessment method for phase {I} clinical trials
  in different ethnic populations.
\newblock {\em Statistics in Medicine}, 34(10):1681--1694, 2015.

\bibitem{petit2018unified}
Caroline Petit, Adeline Samson, Satoshi Morita, Moreno Ursino, J{\'e}r{\'e}mie
  Guedj, Vincent Jullien, Emmanuelle Comets, and Sarah Zohar.
\newblock Unified approach for extrapolation and bridging of adult information
  in early-phase dose-finding paediatric studies.
\newblock {\em Statistical methods in medical research}, 27(6):1860--1877,
  2018.

\bibitem{morita2011application}
Satoshi Morita.
\newblock Application of the continual reassessment method to a phase {I}
  dose-finding trial in {J}apanese patients: East meets west.
\newblock {\em Statistics in Medicine}, 30(17):2090--2097, 2011.

\bibitem{liyuan2019}
Yimei Li and Ying Yuan.
\newblock {PA-CRM}: A continuous reassessment method for pediatric phase {I}
  oncology trials with concurrent adult trials.
\newblock {\em Biometrics}, pages 1--10, 2020.

\bibitem{o1990continual}
John O'Quigley, Margaret Pepe, and Lloyd Fisher.
\newblock Continual reassessment method: a practical design for phase 1
  clinical trials in cancer.
\newblock {\em Biometrics}, pages 33--48, 1990.

\bibitem{babb1998cancer}
James Babb, Andr{\'e} Rogatko, and Shelemyahu Zacks.
\newblock Cancer phase {I} clinical trials: efficient dose escalation with
  overdose control.
\newblock {\em Statistics in Medicine}, 17(10):1103--1120, 1998.

\bibitem{neuenschwander2008critical}
Beat Neuenschwander, Michael Branson, and Thomas Gsponer.
\newblock Critical aspects of the bayesian approach to phase {I} cancer trials.
\newblock {\em Statistics in Medicine}, 27(13):2420--2439, 2008.

\bibitem{yin2009bayesian}
Guosheng Yin and Ying Yuan.
\newblock Bayesian model averaging continual reassessment method in phase {I}
  clinical trials.
\newblock {\em Journal of the American Statistical Association},
  104(487):954--968, 2009.

\bibitem{liu2015bayesian}
Suyu Liu and Ying Yuan.
\newblock Bayesian optimal interval designs for phase {I} clinical trials.
\newblock {\em Journal of the Royal Statistical Society: Series C (Applied
  Statistics)}, 64(3):507--523, 2015.

\bibitem{ji2010modified}
Yuan Ji, Ping Liu, Yisheng Li, and B~Nebiyou~Bekele.
\newblock A modified toxicity probability interval method for dose-finding
  trials.
\newblock {\em Clinical Trials}, 7(6):653--663, 2010.

\bibitem{yan2017keyboard}
Fangrong Yan, Sumithra~J Mandrekar, and Ying Yuan.
\newblock Keyboard: a novel bayesian toxicity probability interval design for
  phase {I} clinical trials.
\newblock {\em Clinical Cancer Research}, 23(15):3994--4003, 2017.

\bibitem{guo2017bayesian}
Wentian Guo, Sue-Jane Wang, Shengjie Yang, Henry Lynn, and Yuan Ji.
\newblock A bayesian interval dose-finding design addressingockham's razor:
  mtpi-2.
\newblock {\em Contemporary Clinical Trials}, 58:23--33, 2017.

\bibitem{yuan2019model}
Ying Yuan, J~Jack Lee, and Susan~G Hilsenbeck.
\newblock Model-assisted designs for early-phase clinical trials: Simplicity
  meets superiority.
\newblock {\em JCO Precision Oncology}, 3:1--12, 2019.

\bibitem{morita2008determining}
Satoshi Morita, Peter~F Thall, and Peter M{\"u}ller.
\newblock Determining the effective sample size of a parametric prior.
\newblock {\em Biometrics}, 64(2):595--602, 2008.

\bibitem{zhou2018comparative}
Heng Zhou, Thomas~A Murray, Haitao Pan, and Ying Yuan.
\newblock Comparative review of novel model-assisted designs for phase {I}
  clinical trials.
\newblock {\em Statistics in Medicine}, 37(14):2208--2222, 2018a.

\bibitem{zhou2018accuracy}
Heng Zhou, Ying Yuan, and Lei Nie.
\newblock Accuracy, safety, and reliability of novel phase {I} trial designs.
\newblock {\em Clinical Cancer Research}, 24(18):4357--4364, 2018b.

\bibitem{clertant2017semiparametric}
Matthieu Clertant and John O’Quigley.
\newblock Semiparametric dose finding methods.
\newblock {\em Journal of the Royal Statistical Society: Series B (Statistical
  Methodology)}, 79(5):1487--1508, 2017.

\end{thebibliography}
\clearpage

\setlength{\tabcolsep}{8pt}
\renewcommand{\arraystretch}{1}
\begin{table}[ht!]
	\centering
	\caption{iBOIN decision boundaries up to 30 patients with a cohort size of 3, given the skeleton $(q_1, 
		\cdots, q_5) = (0.10, 0.19, 0.30, 0.42, 0.54)$ and PESS $n_{01}=\cdots=n_{05} = 3$. The target DLT probability $\phi=0.3$.}
	\begin{tabular}{clcccccccccc}
		\hline
		&       & \multicolumn{10}{c}{Number of patients treated at current dose} \\
		\cline{3-12}
			Dose level & \multicolumn{1}{c}{Action$^*$} & 3     & 6     & 9     & 12    & 15    & 18    & 21    & 24    & 27    & 30 \\
			\hline
		\multirow{2}[0]{*}{1} & {\small Escalate if no. of DLT $\leq$} & 1     & 1     & 2     & 3     & 4     & 4     & 5     & 6     & 6     & 7 \\
		& {\small De-escalate if no. of DLT $\geq$} & 2     & 3     & 4     & 5     & 7     & 8     & 9     & 10    & 11    & 12 \\
			\hline
		\multirow{2}[0]{*}{2} & {\small Escalate if no. of DLT $\leq$} & 0     & 1     & 2     & 3     & 3     & 4     & 5     & 5     & 6     & 7 \\
		& {\small De-escalate if no. of DLT  $\geq$} & 2     & 3     & 4     & 5     & 6     & 7     & 8     & 9     & 11    & 12 \\
			\hline
		\multirow{2}[0]{*}{3} & {\small Escalate if no. of DLT $\leq$} & 0     & 1     & 2     & 2     & 3     & 4     & 4     & 5     & 6     & 7 \\
		& {\small De-escalate if no. of DLT $\geq$ }& 2     & 3     & 4     & 5     & 6     & 7     & 8     & 9     & 10    & 11 \\
			\hline
		\multirow{2}[0]{*}{4} & {\small Escalate if no. of DLT$\leq$ } & 0     & 1     & 1     & 2     & 3     & 3     & 4     & 5     & 6     & 6 \\
		& {\small De-escalate if no. of DLT  $\geq$} & 1     & 2     & 3     & 4     & 6     & 7     & 8     & 9     & 10    & 11 \\
			\hline
		\multirow{2}[0]{*}{5} & {\small Escalate if no. of DLT  $\leq$} & 0     & 0     & 1     & 2     & 2     & 3     & 4     & 5     & 5     & 6 \\
		& {\small De-escalate if no. of DLT  $\geq$} & 1     & 2     & 3     & 4     & 5     & 6     & 7     & 8     & 10    & 11 \\
		\hline
			\multicolumn{12}{l}{ *When neither ``Escalate" nor  ``De-escalate" is triggered, stay at the current dose  for } \\
		\multicolumn{12}{l}{   treating the next  cohort of patients. }
	\end{tabular}%

	\label{tab:boiny}
\end{table}%

\setlength{\tabcolsep}{5pt}
\LTcapwidth=\textwidth
\renewcommand{\arraystretch}{1}
\begin{table}[htbp]
	\centering
	\caption{ Ten dose-toxicity scenarios with target DLT probability $\phi=0.30$. The prior MTDs are correctly specified in scenarios 1-5 and misspecified in scenarios 6-10.}
	\begin{tabular}{lcccccccccccc}
		\hline \hline
		  & \multicolumn{5}{c}{Dose level}        &       & & \multicolumn{5}{c}{Dose level} \\
		  \cline{2-6} \cline{9-13}
		& 1     & 2     & 3     & 4     & 5     & &      & 1     & 2     & 3     & 4     & 5 \\
		\hline\hline
			&&&&&&&&&&&&\\
	 & \multicolumn{5}{c}{\underline{Scenario 1}} &       && \multicolumn{5}{c}{\underline{Scenario 6}} \\
	True  Pr(DLT) & \textit{\textbf{0.30}} & 0.42  & 0.50  & 0.60  & 0.65  &  &     & 0.09  & 0.12  & 0.15  & \textit{\textbf{0.30}} & 0.45 \\
	Prior Pr(DLT) & \textit{\textbf{0.30}} & 0.42  & 0.54  & 0.64  & 0.73  &  &     & 0.01  & 0.04  & 0.10  & 0.19  & \textit{\textbf{0.30}} \\
		&&&&&&&&&&&&\\
	& \multicolumn{5}{c}{\underline{Scenario 2}} &       && \multicolumn{5}{c}{\underline{Scenario 7}} \\
	True  Pr(DLT) & 0.15  & \textit{\textbf{0.27}} & 0.40  & 0.50  & 0.65  & &      & 0.08  & 0.15  & \textit{\textbf{0.31}} & 0.45  & 0.55 \\
	Prior Pr(DLT) & 0.19  & \textit{\textbf{0.30}} & 0.42  & 0.54  & 0.64  & &      & 0.19  & \textit{\textbf{0.30}} & 0.42  & 0.54  & 0.64 \\
		&&&&&&&&&&&&\\
	& \multicolumn{5}{c}{\underline{Scenario 3}}&       && \multicolumn{5}{c}{\underline{Scenario 8}} \\
	True  Pr(DLT) & 0.08  & 0.15  & \textit{\textbf{0.31}} & 0.45  & 0.55  & &      & 0.08  & 0.15  & \textit{\textbf{0.31}} & 0.45  & 0.55 \\
	Prior Pr(DLT) & 0.10  & 0.19  & \textit{\textbf{0.30}} & 0.42  & 0.54  & &      & 0.01  & 0.04  & 0.10  & 0.19  & \textit{\textbf{0.30}} \\
		&&&&&&&&&&&&\\
	& \multicolumn{5}{c}{\underline{Scenario 4}} &       && \multicolumn{5}{c}{\underline{Scenario 9}} \\ 

	True  Pr(DLT) & 0.09  & 0.12  & 0.15  & \textit{\textbf{0.30}} & 0.45  &&       & 0.04  & 0.08  & 0.10  & 0.18  & \textit{\textbf{0.27}} \\
	Prior Pr(DLT) & 0.04  & 0.10  & 0.19  & \textit{\textbf{0.30}} & 0.42  & &      & 0.04  & 0.09  & \textit{\textbf{0.30}} & 0.40  & 0.45 \\
	
			&&&&&&&&&&&&\\
	& \multicolumn{5}{c}{\underline{Scenario 5}} &       && \multicolumn{5}{c}{\underline{Scenario 10}} \\
	True  Pr(DLT) & 0.05  & 0.08  & 0.10  & 0.14  & \textit{\textbf{0.30}} &&       & 0.08  & 0.10  & \textit{\textbf{0.28}} & 0.40  & 0.45 \\
	Prior Pr(DLT) & 0.01  & 0.04  & 0.10  & 0.19  & \textit{\textbf{0.30}} & &      & \textit{\textbf{0.30}} & 0.42  & 0.54  & 0.64  & 0.73 \\	\hline
	\end{tabular}%
	\label{scenarios}
\end{table}%

\clearpage
\setlength{\tabcolsep}{10pt}
\LTcapwidth=\textwidth
\renewcommand{\arraystretch}{0.6}
\begin{longtable}{lcccccc}
	\caption{\sspace Operating characteristics of iCRM,  iBOIN and iKeyboard, in comparison with their counterparts with non-informative priors.  iBOIN$_R$ and iKeyboard$_R$ are iBOIN and iKeyboard using robust priors.}
	\label{OC_target30_summary}\\

	\hline  \\

	&  & \% Patients & \% Patients& Risk of & Risk of  \\
	
	Design & PCS &   at MTD   & above MTD  &overdosing  & poor allocation \\

\hline\\
	
	\endfirsthead
	\caption* {\textbf{Table \ref{OC_target30_summary} Continued:}}\\
	\hline\\

	& \% Correct & \% Patients  & \% Patients& Overdose & \% Poor  \\
	
	Design & selection &   at MTD  & above MTD  &(\%)  & allocation \\
	\hline\\
	\endhead
 & \multicolumn{5}{c}{\textbf{Scenario 1}} \\
     CRM   & 54.8  & 59.9  & 27.8  & 23.2  & 12.2 \\
 iCRM  & 63.1  & 65.2  & 24.9  & 19.4  & 9.8 \\ \rowcolor{Gray}
 BOIN  & 59.2  & 59.6  & 29.0  & 23.6  & 10.2 \\ \rowcolor{Gray}
 iBOIN & 64.2  & 66.2  & 22.4  & 12.8  & 4.5 \\ \rowcolor{Gray}
 iBOIN$_R$ & 64.2  & 66.2  & 22.4  & 12.8  & 4.5 \\
 Keyboard & 59.2  & 59.3  & 29.3  & 23.6  & 10.2 \\
 iKeyboard & 64.2  & 50.7  & 39.6  & 34.2  & 17.8 \\
 iKeyboard$_R$ & 64.2  & 50.7  & 39.6  & 34.2  & 17.8 \\
 &       &       &       &       &  \\
 & \multicolumn{5}{c}{\textbf{Scenario 2}} \\
 CRM   & 51.6  & 36.1  & 7.7   & 29.5  & 25.2 \\
 iCRM  & 53.3  & 42.4  & 5.6   & 23.7  & 16.8 \\ \rowcolor{Gray}
 BOIN  & 50.6  & 41.1  & 6.0   & 23.0  & 17.1 \\ \rowcolor{Gray}
 iBOIN & 57.8  & 47.6  & 3.7   & 10.4  & 8.6 \\ \rowcolor{Gray}
 iBOIN$_R$ & 57.8  & 47.6  & 3.7   & 10.4  & 8.6 \\ 
 Keyboard & 50.2  & 41.1  & 6.0   & 23.0  & 16.7 \\
 iKeyboard & 59.6  & 37.8  & 6.6   & 35.1  & 23.6 \\
 iKeyboard$_R$ & 59.6  & 37.8  & 6.6   & 35.1  & 23.6 \\
 &       &       &       &       &  \\
 & \multicolumn{5}{c}{\textbf{Scenario 3}} \\
 CRM   & 57.2  & 37.8  & 22.1  & 17.3  & 21.6 \\
 iCRM  & 60.2  & 40.4  & 20.9  & 15.3  & 18.8 \\ \rowcolor{Gray}
 BOIN  & 52.3  & 35.6  & 17.0  & 7.9   & 19.2 \\ \rowcolor{Gray}
 iBOIN & 59.8  & 41.3  & 14.5  & 3.5   & 10.9 \\ \rowcolor{Gray} 
 iBOIN$_R$ & 58.9  & 38.2  & 17.7  & 9.1   & 15.2 \\ 
 Keyboard & 52.4  & 35.7  & 17.1  & 7.9   & 18.9 \\
 iKeyboard & 62.5  & 35.8  & 28.8  & 18.7  & 19.1 \\
 iKeyboard$_R$ & 59.7  & 35.6  & 29.0  & 19.4  & 19.7 \\
 &       &       &       &       &  \\
 & \multicolumn{5}{c}{\textbf{Scenario 4}} \\
 CRM   & 52.0  & 30.0  & 15.3  & 10.3  & 33.4 \\
 iCRM  & 56.6  & 33.8  & 14.4  & 8.6   & 26.5 \\  \rowcolor{Gray}
 BOIN  & 51.5  & 28.6  & 13.1  & 1.2   & 24.6 \\ \rowcolor{Gray}
 iBOIN & 59.7  & 36.0  & 12.1  & 0.6   & 12.8 \\  \rowcolor{Gray}
 iBOIN$_R$ & 57.6  & 32.4  & 15.7  & 3.2   & 19.0 \\
 Keyboard & 52.1  & 28.6  & 13.1  & 1.2   & 24.6 \\
 iKeyboard & 65.1  & 31.7  & 25.6  & 11.2  & 18.9 \\
 iKeyboard$_R$ & 62.4  & 31.7  & 25.6  & 11.2  & 18.9 \\
 &       &       &       &       &  \\
 & \multicolumn{5}{c}{\textbf{Scenario 5}} \\
 CRM   & 72.7  & 38.6  & 0     & 0     & 23.4 \\
 iCRM  & 75.8  & 41.7  & 0     & 0     & 19.7 \\ \rowcolor{Gray}
 BOIN  & 71.0  & 35.2  & 0     & 0     & 16.8 \\ \rowcolor{Gray}
 iBOIN & 76.8  & 42.2  & 0     & 0     & 9.6 \\ \rowcolor{Gray}
 iBOIN$_R$ & 76.8  & 42.2  & 0     & 0     & 9.6 \\
 Keyboard & 71.0  & 35.2  & 0     & 0     & 16.8 \\
 iKeyboard & 75.8  & 47.1  & 0     & 0     & 5.2 \\
 iKeyboard$_R$ & 75.8  & 47.1  & 0     & 0     & 5.2 \\
 &       &       &       &       &  \\
 & \multicolumn{5}{c}{\textbf{Scenario 6}} \\
 CRM   & 50.7  & 29.9  & 14.6  & 9.3   & 32.9 \\
 iCRM  & 57.3  & 33.8  & 17.0  & 13.0  & 28.2 \\ \rowcolor{Gray}
 BOIN  & 51.5  & 28.6  & 13.1  & 1.2   & 24.6 \\ \rowcolor{Gray}
 iBOIN & 58.6  & 35.5  & 18.4  & 3.8   & 11.8 \\ \rowcolor{Gray}
 iBOIN$_R$ & 58.6  & 35.5  & 18.4  & 3.8   & 11.8 \\
 Keyboard & 52.1  & 8.6   & 13.1  & 1.2   & 24.6 \\
 iKeyboard & 59.5  & 9.2   & 27.9  & 11.2  & 17.9 \\
 iKeyboard$_R$ & 59.5  & 9.2   & 27.9  & 11.2  & 17.9 \\
 &       &       &       &       &  \\
 & \multicolumn{5}{c}{\textbf{Scenario 7}} \\
 CRM   & 58.0  & 38.1  & 21.7  & 17.3  & 21.8 \\
 iCRM  & 59.8  & 38.0  & 18.3  & 13.4  & 21.3 \\ \rowcolor{Gray}
 BOIN  & 52.3  & 35.6  & 17.0  & 7.9   & 19.2 \\ \rowcolor{Gray}
 iBOIN & 61.6  & 33.0  & 10.9  & 2.2   & 14.8 \\ \rowcolor{Gray}
 iBOIN$_R$ & 61.6  & 33.0  & 10.9  & 2.2   & 14.8 \\
 Keyboard & 52.4  & 35.7  & 17.1  & 7.9   & 18.9 \\
 iKeyboard & 56.4  & 45.4  & 15.3  & 3.6   & 6.8 \\
 iKeyboard$_R$ & 56.4  & 45.4  & 15.3  & 3.6   & 6.8 \\
 &       &       &       &       &  \\
 & \multicolumn{5}{c}{\textbf{Scenario 8}} \\
 CRM   & 57.8  & 37.0  & 21.6  & 17.2  & 22.9 \\
 iCRM  & 58.7  & 41.5  & 25.5  & 21.3  & 20.3 \\ \rowcolor{Gray}
 BOIN  & 52.3  & 35.6  & 17.0  & 7.9   & 19.2 \\ \rowcolor{Gray}
 iBOIN & 54.3  & 36.2  & 28.6  & 19.9  & 19.8 \\ \rowcolor{Gray}
 iBOIN$_R$ & 54.3  & 36.2  & 28.6  & 19.9  & 19.8 \\
 Keyboard & 52.4  & 35.7  & 17.1  & 7.9   & 18.9 \\
 iKeyboard & 46.8  & 33.0  & 37.8  & 34.4  & 25.2 \\
 iKeyboard$_R$ & 46.8  & 33.0  & 37.8  & 34.4  & 25.2 \\
 
 &       &       &       &       &  \\
 & \multicolumn{5}{c}{\textbf{Scenario 9}} \\
 CRM   & 67.5  & 36.5  & 0     & 0     & 30.0 \\
 iCRM  & 64.8  & 35.3  & 0     & 0     & 31.8 \\ \rowcolor{Gray}
 BOIN  & 69.4  & 33.8  & 0     & 0     & 22.4 \\ \rowcolor{Gray}
 iBOIN & 51.4  & 25.7  & 0     & 0     & 35.3 \\ \rowcolor{Gray}
 iBOIN$_R$ & 68.8  & 36.7  & 0     & 0     & 21.2 \\
 Keyboard & 69.4  & 33.8  & 0     & 0     & 22.4 \\
 iKeyboard & 58.7  & 39.8  & 0     & 0     & 17.8 \\
 iKeyboard$_R$ & 71.7  & 39.8  & 0     & 0     & 17.8 \\
 &       &       &       &       &  \\
 & \multicolumn{5}{c}{\textbf{Scenario 10}} \\
 CRM   & 54.2  & 36.5  & 0.0   & 28.3  & 26.1 \\
 iCRM  & 60.6  & 40.0  & 0.0   & 15.2  & 18.1 \\ \rowcolor{Gray}
 BOIN  & 53.1  & 37.5  & 5.3   & 14.6  & 17.2 \\ \rowcolor{Gray}
 iBOIN & 65.5  & 36.1  & 0.5   & 3.1   & 13.4 \\ \rowcolor{Gray}
 iBOIN$_R$ & 65.5  & 36.1  & 0.5   & 3.1   & 13.4 \\
 Keyboard & 52.6  & 37.5  & 5.3   & 14.6  & 17.1 \\
 iKeyboard & 66.5  & 37.9  & 1.6   & 3.8   & 7.0 \\
 iKeyboard$_R$ & 66.5  & 37.9  & 1.6   & 3.8   & 7.0 \\
\hline

\end{longtable}

\clearpage

\clearpage
\pagenumbering{arabic}
\renewcommand*{\thepage}{A.\arabic{page}}
\setcounter{table}{0}
\renewcommand{\thetable}{A.\arabic{table}}
\setcounter{section}{0}
\renewcommand*{\thesection}{A.\arabic{section}}
\setcounter{figure}{0}
\renewcommand{\thefigure}{A.\arabic{figure}}

\begin{huge}
	\begin{center}\textbf{Appendix}
	\end{center}
\end{huge}

\section{Determining informative prior for BOIN} \label{sec:derivation}
Suppose at dose level $j$, the prior estimate of DLT probability is $q_j$ with PESS of $n_0$.  This prior information can be transformed into the prior distribution of the three hypothesis employed by BOIN (i.e., $H_{1j}: p_j = \phi,  \quad H_{2j}: p_j = \phi_1, \quad H_{3j}: p_j = \phi_2$) as follows: for $k=1$, 2 and 3, 
\begin{align}
\begin{split}
\pi_{kj}&=\mbox{Pr}(H_{kj}\mid n_0,q_j) \\
&=\sum_{x=0}^{n_0}\mbox{Pr}(H_{kj}\mid x)\mbox{Pr}(x\mid n_0,q_j)\\
&=\sum_{x=0}^{n_0}\frac{\mbox{Pr}\left(x\mid H_{kj}\right)\mbox{Pr}\left(H_{kj}\right)}{\sum_{k'=1}^{3}\mbox{Pr}\left(x\mid H_{kj}\right)\mbox{Pr}\left(H_{kj}\right)}\mbox{Pr}(x\mid n_0,q_j)\\
&=\sum_{x=0}^{n_0}\frac{\mbox{Pr}\left(x\mid H_{kj}\right)}{\sum_{k'=1}^{3}\mbox{Pr}\left(x\mid H_{kj}\right)}\mbox{Pr}(x\mid n_0,q_j).\\
&=\sum_{x=0}^{n_0}\frac{\phi_k^x(1-\phi_k)^{n_0-x}}{\sum_{k'=1}^{3}\phi_k'^x(1-\phi_{k'})^{n_0-x}}\binom{n_0}{x}q_j^x(1-q_j)^{n_0-x}.\\
\end{split}
\label{postpi}
\end{align}
By doing so, the prior information is incorporated into the dose escalation and de-escalation boundaries, as given by equation (2.3). 
\clearpage
\section{iBOIN Shiny app interface}
\begin{figure}[h]
	\begin{center}
		\includegraphics[scale=0.8]{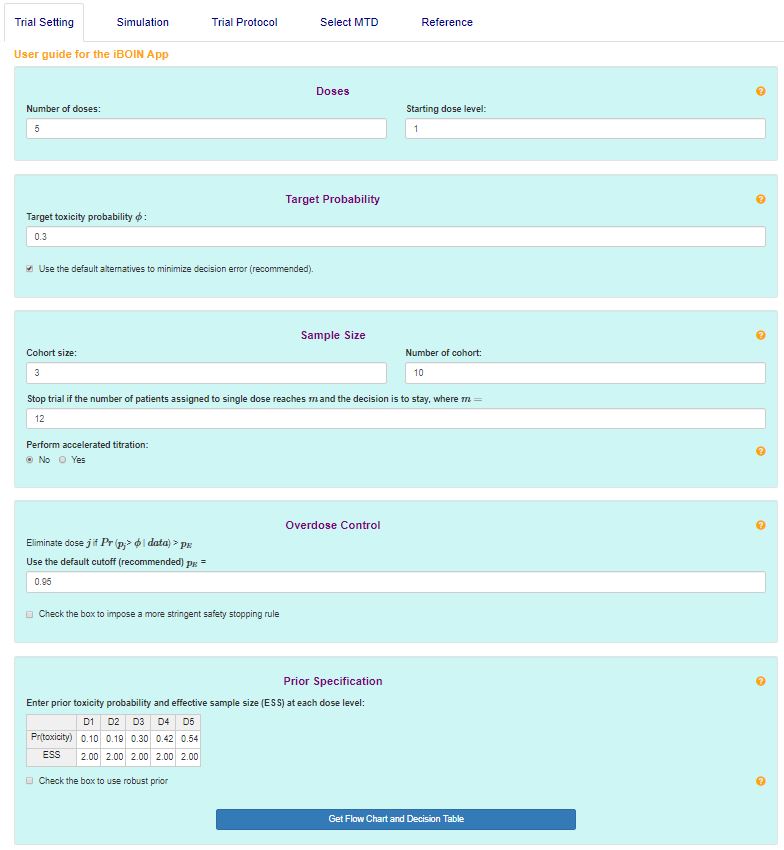}
		\caption{User interface of iBOIN software.}
		\label{fig:app}
		
	\end{center}
\end{figure}

\clearpage
\section{Random scenario configuration}\label{sec:generate_random}
\subsection{Generate random scenarios where prior MTD is correctly specified}
To examine how the informative designs perform, we generated 2000 random scenarios with the MTD located at dose level 1, 2, 3, 4, and 5, with equal probability. The random scenarios were generated using the following pseudo-uniform algorithm \cite{clertant2017semiparametric}.  Given a target DLT probability $\phi$ and $J$ dose levels,
\begin{enumerate}
	\item Select one of the $j\in (1,\cdots, J)$ with probability $1/J$.
	\item  Sample $M\sim\mbox{Beta}(\mbox{max}\{J-j,0.5\},1)$.
	\item Repeatedly sample $J$ toxicity probabilities uniformly on $[0, B]$ until these correspond to a scenario in which dose level $j$ is the MTD, where $B=\phi+(1-\phi)\times M$ is the upper bound of DLT probability.	
\end{enumerate}

\begin{figure}[h]
	\begin{center}
	\includegraphics[scale=0.55]{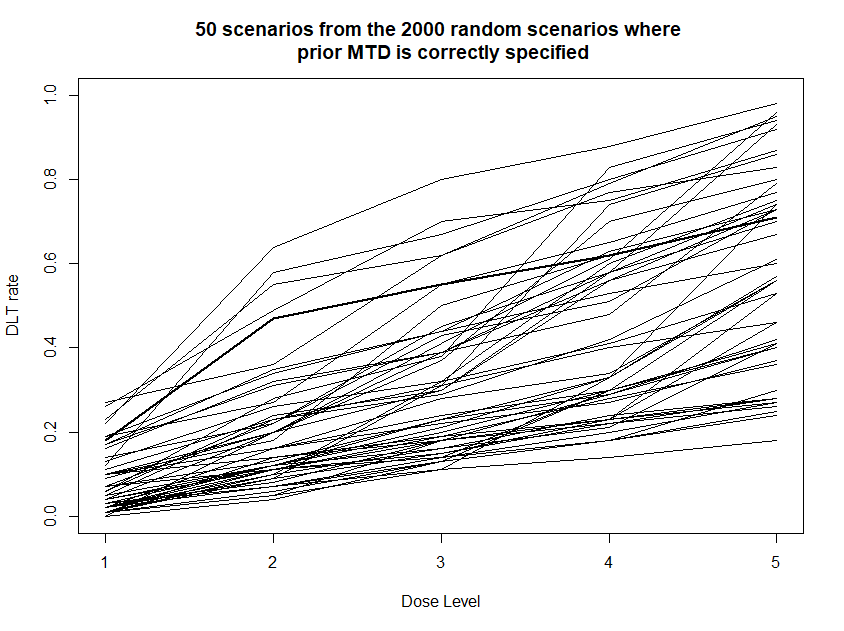}
	\caption{50 randomly selected scenarios from the 2000 scenarios generated}
	\label{fig:rand_scen}
	\end{center}
\end{figure}

In these scenarios, the MTD is the dose with the DLT probability closest, but not necessarily equal to the target $\phi$. It is possible to obtain scenarios in which all of the doses have DLT probabilities below or above the target $\phi$. To ensure that MTD is uniquely and meaningfully defined, we required that the true DLT probability of the MTD be within $[\phi-0.05, \phi+0.05]$, and the distance between the MTD and its adjacent doses be greater than 0.05 and less than 0.3, i.e., $0.05<p_{j+1}-p_j<0.3$ and $0.05<p_j-p_{j-1} < 0.3$. The generating process was stopped until we obtained 2000 random scenarios that satisfied the specification. Figure \ref{fig:rand_scen} shows 50 scenarios from the 2000 random scenarios generated. The plot shows that the scenarios cover a wide range of possible dose-toxicity scenarios that we may encounter in practice. Each of the 2000 scenarios has their prior MTD correctly specified. The five prior skeletons used are presented in scenarios 1-5 in Table \ref{scenarios}.

\clearpage

\subsection{Generate random scenarios with different levels of misspecification}\label{sec:fivesets}
To assess the performance of the informative designs when the prior is misspecified (i.e., the prior MTD is not corresponding to the true MTD), we conducted extensive simulation for random scenarios with different levels of severity for mis-specification. Below are the configurations of the scenarios.  
\begin{enumerate}
	
	\item \textbf{The prior MTD is one dose below the true MTD}. Generate 2000 scenarios with true MTD located at dose level 2, 3, 4, and 5 with equal probability. The corresponding prior skeletons are in scenarios 1, 2, 3, and 4 in Table \ref{scenarios}. 
	\item \textbf{The prior MTD is one dose above the MTD}. Generate 2000 scenarios with true MTD located at dose level 1, 2, 3, and 4 with equal probability. The corresponding prior skeletons are in scenarios 2, 3, 4, and 5 in Table \ref{scenarios}. 
	\item \textbf{The prior MTD is two doses below the true MTD}. Generate 2000 scenarios with true MTD located at dose level 3, 4, and 5 with equal probability. The corresponding prior skeletons are in scenarios 1, 2, and 3 in Table \ref{scenarios}. 
	\item \textbf{The prior MTD is two doses above the MTD}. Generate 2000 scenarios with true MTD located at dose level 1, 2, and 3 with equal probability. The corresponding prior skeletons are in scenarios 3, 4, and 5 in Table \ref{scenarios}.
\end{enumerate}

\clearpage
\section{Simulation results for mixture prior under 10 scenarios in Table 2. }\label{sec:mixture}

Table \ref{OC_target30_summary_mix} shows the performance of the iBOIN and iKeyboard designs when the mixture prior was used. iBOIN$_{M50}$ and iBOIN$_{M90}$ denote iBOIN using mixture prior when the weight assigned to the informative prior component is $w=0.5$ and $w=0.9$, respectively.  iKeyboard$_{M50}$ and iKeyboard$_{M90}$ are defined similarly. For ease of comparison, the results of BOIN, iBOIN and iBOIN$_R$ are replicated from Table 3.  The results show that in general mixture prior (e.g., iBOIN$_{M50}$ and iBOIN$_{M90}$) does not perform as well as the robust prior (iBOIN$_R$). 


\setlength{\tabcolsep}{10pt}
\LTcapwidth=\textwidth
\renewcommand{\arraystretch}{0.6}
\begin{longtable}{lcccccc}
	\caption{\sspace Operating characteristics of iBOIN and iKeyboard designs with the mixture priors.  }
	\label{OC_target30_summary_mix}\\

	\hline  \\
	
	&  & \% Patients & \% Patients& Risk of & Risk of  \\
	
	Design & PCS &   at MTD   & above MTD  &overdosing  & poor allocation \\

	\hline\\
	
	\endfirsthead
	\caption* {\textbf{Table \ref{OC_target30_summary_mix} Continued:}}\\
	\hline\\
	
	& \% Correct & \% Patients  & \% Patients& Overdose & \% Poor  \\
	
	Design & selection &   at MTD  & above MTD  &(\%)  & allocation \\
	\hline\\
	\endhead
	& \multicolumn{5}{c}{\textbf{Scenario 1}} \\ \rowcolor{Gray}
	BOIN  & 59.2  & 59.6  & 29.0  & 23.6  & 10.2 \\ \rowcolor{Gray}
	iBOIN & 64.2  & 66.2  & 22.4  & 12.8  & 4.5 \\ \rowcolor{Gray}
	iBOIN$_R$ & 64.2  & 66.2  & 22.4  & 12.8  & 4.5 \\ \rowcolor{Gray}
	iBOIN$_{M50}$ & 61.6  & 63.2  & 25.1  & 17.8  & 6.9 \\ \rowcolor{Gray}
	iBOIN$_{M90}$ & 63.2  & 65.8  & 22.5  & 13.1  & 5.3 \\
	Keyboard & 59.2  & 59.3  & 29.3  & 23.6  & 10.2 \\
	iKeyboard & 64.2  & 50.7  & 39.6  & 34.2  & 17.8 \\
	iKeyboard$_R$ & 64.2  & 50.7  & 39.6  & 34.2  & 17.8 \\
	iKeyboard$_{M50}$ & 59.1  & 54.2  & 34.9  & 29.4  & 13.5 \\
	iKeyboard$_{M90}$ & 63.8  & 51.0  & 38.9  & 32.2  & 16.7 \\
	&       &       &       &       &  \\
	& \multicolumn{5}{c}{\textbf{Scenario 2}} \\ \rowcolor{Gray}
	BOIN  & 50.6  & 41.1  & 6.0   & 23.0  & 17.1 \\ \rowcolor{Gray}
	iBOIN & 57.8  & 47.6  & 3.7   & 10.4  & 8.6 \\ \rowcolor{Gray}
	iBOIN$_R$ & 57.8  & 47.6  & 3.7   & 10.4  & 8.6 \\ \rowcolor{Gray}
	iBOIN$_{M50}$ & 53.8  & 44.4  & 4.8   & 16.2  & 11.3 \\ \rowcolor{Gray}
	iBOIN$_{M90}$ & 58.1  & 47.1  & 3.9   & 11.8  & 8.8 \\
	Keyboard & 50.2  & 41.1  & 6.0   & 23.0  & 16.7 \\
	iKeyboard & 59.6  & 37.8  & 6.6   & 35.1  & 23.6 \\
	iKeyboard$_R$ & 59.6  & 37.8  & 6.6   & 35.1  & 23.6 \\
	iKeyboard$_{M50}$ & 52.5  & 39.2  & 7.1   & 29.3  & 20.7 \\
	iKeyboard$_{M90}$ & 57.9  & 38.2  & 7.0   & 33.3  & 22.9 \\
	&       &       &       &       &  \\
	& \multicolumn{5}{c}{\textbf{Scenario 3}} \\ \rowcolor{Gray}
	BOIN  & 52.3  & 35.6  & 17.0  & 7.9   & 19.2 \\ \rowcolor{Gray}
	iBOIN & 59.8  & 41.3  & 14.5  & 3.5   & 10.9 \\ \rowcolor{Gray}
	iBOIN$_R$ & 58.9  & 38.2  & 17.7  & 9.1   & 15.2 \\ \rowcolor{Gray}
	iBOIN$_{M50}$ & 57.0  & 39.7  & 14.7  & 4.0   & 13.9 \\ \rowcolor{Gray}
	iBOIN$_{M90}$ & 61.3  & 42.1  & 13.7  & 2.9   & 11.1 \\
	Keyboard & 52.4  & 35.7  & 17.1  & 7.9   & 18.9 \\
	iKeyboard & 62.5  & 35.8  & 28.8  & 18.7  & 19.1 \\
	iKeyboard$_R$ & 59.7  & 35.6  & 29.0  & 19.4  & 19.7 \\
	iKeyboard$_{M50}$ & 55.5  & 36.4  & 22.6  & 12.3  & 17.5 \\
	iKeyboard$_{M90}$ & 62.0  & 36.2  & 26.8  & 16.1  & 19.3 \\
	&       &       &       &       &  \\
	& \multicolumn{5}{c}{\textbf{Scenario 4}} \\ \rowcolor{Gray}
	BOIN  & 51.5  & 28.6  & 13.1  & 1.2   & 24.6 \\ \rowcolor{Gray}
	iBOIN & 59.7  & 36.0  & 12.1  & 0.6   & 12.8 \\ \rowcolor{Gray}
	iBOIN$_R$ & 57.6  & 32.4  & 15.7  & 3.2   & 19.0 \\ \rowcolor{Gray}
	iBOIN$_{M50}$ & 55.8  & 32.4  & 12.8  & 1.1   & 17.5 \\ \rowcolor{Gray}
	iBOIN$_{M90}$ & 60.4  & 35.8  & 12.2  & 0.9   & 12.9 \\
	Keyboard & 52.1  & 28.6  & 13.1  & 1.2   & 24.6 \\
	iKeyboard & 65.1  & 31.7  & 25.6  & 11.2  & 18.9 \\
	iKeyboard$_R$ & 62.4  & 31.7  & 25.6  & 11.2  & 18.9 \\
	iKeyboard$_{M50}$ & 56.5  & 30.4  & 19.4  & 5.2   & 21.6 \\
	iKeyboard$_{M90}$ & 64.0  & 31.4  & 24.4  & 9.7   & 19.6 \\
	&       &       &       &       &  \\
	& \multicolumn{5}{c}{\textbf{Scenario 5}} \\ \rowcolor{Gray}
	BOIN  & 71.0  & 35.2  & 0     & 0     & 16.8 \\ \rowcolor{Gray}
	iBOIN & 76.8  & 42.2  & 0     & 0     & 9.6 \\ \rowcolor{Gray}
	iBOIN$_R$ & 76.8  & 42.2  & 0     & 0     & 9.6 \\ \rowcolor{Gray}
	iBOIN$_{M50}$ & 72.8  & 38.0  & 0     & 0     & 12.8 \\ \rowcolor{Gray}
	iBOIN$_{M90}$ & 75.2  & 40.3  & 0     & 0     & 11.0 \\
	Keyboard & 71.0  & 35.2  & 0     & 0     & 16.8 \\
	iKeyboard & 75.8  & 47.1  & 0     & 0     & 5.2 \\
	iKeyboard$_R$ & 75.8  & 47.1  & 0     & 0     & 5.2 \\
	iKeyboard$_{M50}$ & 73.2  & 41.7  & 0     & 0     & 8.9 \\
	iKeyboard$_{M90}$ & 76.8  & 46.1  & 0     & 0     & 5.7 \\
	&       &       &       &       &  \\
	& \multicolumn{5}{c}{\textbf{Scenario 6}} \\ \rowcolor{Gray}
	BOIN  & 51.5  & 28.6  & 13.1  & 1.2   & 24.6 \\ \rowcolor{Gray}
	iBOIN & 58.6  & 35.5  & 18.4  & 3.8   & 11.8 \\ \rowcolor{Gray}
	iBOIN$_R$ & 58.6  & 35.5  & 18.4  & 3.8   & 11.8 \\ \rowcolor{Gray}
	iBOIN$_{M50}$ & 55.6  & 32.3  & 16.1  & 3.2   & 17.9 \\ \rowcolor{Gray}
	iBOIN$_{M90}$ & 57.5  & 35.0  & 17.9  & 4.5   & 14.0 \\
	Keyboard & 52.1  & 8.6   & 13.1  & 1.2   & 24.6 \\
	iKeyboard & 59.5  & 9.2   & 27.9  & 11.2  & 17.9 \\
	iKeyboard$_R$ & 59.5  & 9.2   & 27.9  & 11.2  & 17.9 \\
	iKeyboard$_{M50}$ & 58.1  & 9.2   & 21.2  & 5.2   & 20.2 \\
	iKeyboard$_{M90}$ & 60.1  & 0.2   & 26.9  & 9.7   & 18.3 \\
	&       &       &       &       &  \\
	& \multicolumn{5}{c}{\textbf{Scenario 7}} \\ \rowcolor{Gray}
	BOIN  & 52.3  & 35.6  & 17.0  & 7.9   & 19.2 \\ \rowcolor{Gray}
	iBOIN & 61.6  & 33.0  & 10.9  & 2.2   & 14.8 \\ \rowcolor{Gray}
	iBOIN$_R$ & 61.6  & 33.0  & 10.9  & 2.2   & 14.8 \\ \rowcolor{Gray}
	iBOIN$_{M50}$ & 59.2  & 35.5  & 12.9  & 3.5   & 15.2 \\ \rowcolor{Gray}
	iBOIN$_{M90}$ & 61.1  & 33.6  & 10.2  & 2.0   & 14.6 \\
	Keyboard & 52.4  & 35.7  & 17.1  & 7.9   & 18.9 \\
	iKeyboard & 56.4  & 45.4  & 15.3  & 3.6   & 6.8 \\
	iKeyboard$_R$ & 56.4  & 45.4  & 15.3  & 3.6   & 6.8 \\
	iKeyboard$_{M50}$ & 57.6  & 40.9  & 15.5  & 4.9   & 10.7 \\
	iKeyboard$_{M90}$ & 57.8  & 44.1  & 15.1  & 4.6   & 8.8 \\
	&       &       &       &       &  \\
	& \multicolumn{5}{c}{\textbf{Scenario 8}} \\ \rowcolor{Gray}
	BOIN  & 52.3  & 35.6  & 17.0  & 7.9   & 19.2 \\ \rowcolor{Gray}
	iBOIN & 54.3  & 36.2  & 28.6  & 19.9  & 19.8 \\ \rowcolor{Gray}
	iBOIN$_R$ & 54.3  & 36.2  & 28.6  & 19.9  & 19.8 \\ \rowcolor{Gray}
	iBOIN$_{M50}$ & 54.8  & 37.0  & 22.6  & 12.2  & 18.1 \\ \rowcolor{Gray}
	iBOIN$_{M90}$ & 55.1  & 36.5  & 27.1  & 16.2  & 19.4 \\
	Keyboard & 52.4  & 35.7  & 17.1  & 7.9   & 18.9 \\
	iKeyboard & 46.8  & 33.0  & 37.8  & 34.4  & 25.2 \\
	iKeyboard$_R$ & 46.8  & 33.0  & 37.8  & 34.4  & 25.2 \\
	iKeyboard$_{M50}$ & 51.9  & 35.3  & 28.0  & 18.1  & 18.2 \\
	iKeyboard$_{M90}$ & 48.0  & 33.7  & 35.2  & 29.8  & 23.8 \\
	&       &       &       &       &  \\
	& \multicolumn{5}{c}{\textbf{Scenario 9}} \\ \rowcolor{Gray}
	BOIN  & 69.4  & 33.8  & 0     & 0     & 22.4 \\ \rowcolor{Gray}
	iBOIN & 51.4  & 25.7  & 0     & 0     & 35.3 \\ \rowcolor{Gray}
	iBOIN$_R$ & 68.8  & 36.7  & 0     & 0     & 21.2 \\ \rowcolor{Gray}
	iBOIN$_{M50}$ & 59.5  & 28.6  & 0     & 0     & 30.7 \\ \rowcolor{Gray}
	iBOIN$_{M90}$ & 51.6  & 25.3  & 0     & 0     & 36.4 \\
	Keyboard & 69.4  & 33.8  & 0     & 0     & 22.4 \\
	iKeyboard & 58.7  & 39.8  & 0     & 0     & 17.8 \\
	iKeyboard$_R$ & 71.7  & 39.8  & 0     & 0     & 17.8 \\
	iKeyboard$_{M50}$ & 66.8  & 36.4  & 0     & 0     & 20.2 \\
	iKeyboard$_{M90}$ & 62.3  & 38.7  & 0     & 0     & 18.3 \\
	&       &       &       &       &  \\
	& \multicolumn{5}{c}{\textbf{Scenario 10}} \\ \rowcolor{Gray}
	BOIN  & 53.1  & 37.5  & 5.3   & 14.6  & 17.2 \\ \rowcolor{Gray}
	iBOIN & 65.5  & 36.1  & 0.5   & 3.1   & 13.4 \\ \rowcolor{Gray}
	iBOIN$_R$ & 65.5  & 36.1  & 0.5   & 3.1   & 13.4 \\ \rowcolor{Gray}
	iBOIN$_{M50}$ & 62.8  & 37.6  & 2.1   & 5.4   & 11.8 \\ \rowcolor{Gray}
	iBOIN$_{M90}$ & 64.9  & 36.9  & 0.6   & 2.6   & 12.3 \\ 
	Keyboard & 52.6  & 37.5  & 5.3   & 14.6  & 17.1 \\
	iKeyboard & 66.5  & 37.9  & 1.6   & 3.8   & 7.0 \\
	iKeyboard$_R$ & 66.5  & 37.9  & 1.6   & 3.8   & 7.0 \\
	iKeyboard$_{M50}$ & 61.7  & 38.0  & 3.2   & 7.1   & 9.6 \\
	iKeyboard$_{M90}$ & 65.8  & 37.6  & 2.2   & 4.5   & 8.6 \\
	\hline
	\label{tab:addlabel}%
\end{longtable}%

\end{document}